\documentclass[journal,9pt]{IEEEtran}
\pdfoutput=1

\usepackage{amsmath,amsfonts}
\usepackage{algorithmic}
\usepackage{algorithm}
\usepackage{array}
\usepackage[justification=centering]{caption}

\usepackage[caption=false,font=normalsize,labelfont=sf,textfont=sf]{subfig}
\usepackage{textcomp}
\usepackage{stfloats}
\usepackage{url}
\usepackage{verbatim}
\usepackage{graphicx}
\usepackage{cite}
\usepackage{xcolor}
\usepackage{ragged2e} 
\usepackage{booktabs,makecell, multirow, tabularx}

\definecolor{ccr}{RGB}{10,110,150}
\usepackage{hyperref}
\hypersetup{
hypertex=true,
colorlinks=true,
linkcolor=blue,
anchorcolor=magenta,
citecolor=cyan
}

\newcommand{\cyanbf}[1]{{\color{magenta}{\bf #1}}}

\newcommand{\cyan}[1]{{\color{magenta}{#1}}}

\begin{document}
\title{A Dual-Polarization Feature Fusion Network for Radar Automatic Target Recognition Based On HRRP Sequences}

\author{
    Yangbo Zhou, Sen Liu, Hong-Wei Gao, Hai lin, Guohua Wei, Xiaoqing Wang, and Xiao-Min Pan~\IEEEmembership{Member,~IEEE}
\thanks{This work was supported by NSFC under Grant 62171033. 
(Corresponding author: Xiao-Min Pan and Hong-Wei Gao)

Y. Zhou, S. Liu, H.-W. Gao, and X.-M. Pan are with School of Cyberspace Science and Technology, Beijing Institute of Technology, Beijing 100081, China. (e-mail: xmpan@bit.edu.cn, gaohwfd@hotmail.com)

G. Wei is with the School of Information and Electronics, Beijing Institute of Technology, Beijing 100081, China. (e-mail: ghwei@bit.edu.cn)

H. lin is with the College of Physical Science and Technology, Central China Normal University, Wuhan 430079, China. (e-mail:linhai@ccnu.edu.cn)

X. Wang is with the School of Electronics and Communication Engineering, Sun Yat-Sen University, Shenzhen 518107, China, and also with the PengCheng Laboratory, Shenzhen 518055, China (e-mail: wangxq58@mail.sysu.edu.cn) 

}
}

\markboth{IEEE Transactions on Antennas and Propagation,~Vol. ~8, No. 23, 2023}
{Shell \MakeLowercase{\textit{et al. }}: A Sample Article Using IEEEtran. cls for IEEE Journals}

\IEEEpubid{0000--0000~\copyright~2023 IEEE}

\maketitle

\begin{abstract}
Recent advances in radar automatic target recognition (RATR) techniques utilizing deep neural networks have demonstrated remarkable performance, largely due to their robust generalization capabilities. 
To address the challenge for applications with polarimetric HRRP sequences, a dual-polarization feature fusion network (DPFFN) is proposed along with a novel two-stage feature fusion strategy. 
Moreover, a specific fusion loss function is developed, which enables the adaptive generation of comprehensive multi-modal representations from polarimetric HRRP sequences.
Experimental results demonstrate that the proposed network significantly improves performance in radar target recognition tasks, thus validating its effectiveness. The PyTorch implementation of our proposed DPFFN is available at \url{https://github.com/xmpan/DPFFN}. 
\end{abstract}


\section{Introduction}\label{sec:intro}

\IEEEPARstart{R}{ADAR} automatic target recognition (RATR) technology, originally developed for military applications, now plays a crucial role in a wide range of civilian and military fields, including aviation safety, maritime surveillance, and disaster response \cite{willis2007advances,ozdemir2021inverse}. 
This technology is particularly valuable due to its capability to perform multi-target remote sensing under all-weather, all-day, and long-distance conditions, making it indispensable for comprehensive situational awareness and target monitoring \cite{high2021zhou,radar2022bai,cGAN2022wang}. 

Existing RATR methods can be divided into three categories, according to the employed type of data: radar cross section (RCS)~\cite{Lee2016rcs,spaceRCS2018chen,cai2021rcs,rcs2023ye,spaceRCS2023kim}, High-Resolution Range Profile (HRRP) sequences~\cite{spaceHRRP2013da,spaceHRRP2019Persico,spaceHRRP2019wang,RNN2021Chen,chen2022target}, and inverse synthetic aperture radar (ISAR) images~\cite{kim2005isar,spaceISAR2015wang,park2015efficient,spaceISAR2018jia,xue2021saisar,xue2022sequential}. 
Generally, among all those types of data, RCS is the easiest to be acquired, which explains its wide usage. 
However, extracting structural characteristics of a target from its RCS data requires skills and an eye for detail. 
Consequently, RCS-based methods are quite challenging in achieving a high accuracy, especially for complicated applications. 
The ISAR imaging plays a vital role in RATR by providing detailed target characteristics and attitude information. 
However, obtaining high quality ISAR images is much more expensive than that of RCS. 
Additionally, the quality of an ISAR image is always limited by the complexity of imaging algorithms, dependence on target motion, and sensitivity to electromagnetic interference and noise~\cite{cao2021ship, jordan-isar}. 
The HRRP consists of the coherent sum of radar returns when the target scatters along the radar line of sight (RLOS), containing structural information of the target's scattering centers and size information in the RLOS direction~\cite{hrrp2013du}. 
Compared with RCS data, HRRP provides the detailed target structural information in a more explicit manner. 
In comparison with ISAR images, HRRP sequences are easier to be acquired, stored, and processed, while still providing the necessary information under a sufficiently high accuracy. 
Additionally, HRRP sequences contain temporal information, which can be used to track changes of the target over time, providing a dynamic perspective for accurate and reliable RATR in different operating environments. 

The exiting HRRP based RATR approaches can be divided into two categories: traditional approaches and that taking advantages of machine learning (ML). 
The first type depends heavily on a mathematical/physical model to describe the scatterings or echoes from the targets. 
Among many models, that based on scattering centers are the commonly employed. 
By traditional signal processing approaches, different kinds of representations/features, e.g., bispectra, higher order spectra, statistical characteristics, phase information and etc., are generated/extracted to realize recognition tasks~\cite{spectra1993hudson,spectra2001zhang,spectra2001xian-da,spectra2005du, du2006two, du2008radar, du2008radar2, du2011bayesian, md2017pan,time-range2015ai}. 
Generally, these methods always rely excessively on expert experience and prior knowledge although they can perform nice. 
The second category of methods exploit the power of ML, which can automatically learn suitable features from the HRRPs to accomplish the recognition task. 
To conduct recognition tasks, many deep learning (DL) architectures have been employed, e.g., CNN, RNN, LSTM, Transformer~\cite{deepnet2018Jia, RNN2019Xu, zhang2021poltransformer, net2017feng, spaceHRRP2023zhang, zheng2023space, liu2024prior}. 
The approaches based on ML can obtain features/information that cannot be extracted by the widely employed mathematical/physical models. 
As a result, they can perform well in RATR tasks which may not be accomplished by the traditional approaches. 

It is well-known that radar echoes of different polarizations contain rich information associated with the target~\cite{huynen1965measurement,cameron1990feature,cloude1997entropy}. 
The polarization characteristic can thus play a crucial role in RATR based on HRRP~\cite{Liu2021Multi-Polarization}. 
Many efforts have been devoted to exploiting polarization characteristics and thus improving the performance of the RATR. 
Similar to the single polarization case, researches on polarimetric HRRP recognition can also be divided into two technical lines: that based on traditional methods and those on ML/DL. 
The key challenge in traditional polarimetric HRRP recognition methods relates in how to extract and represent polarization features in a highly discriminative form and thus design the related classifiers~\cite{pol2015sheng, pol2019yin}. 
With the rapid advancement of ML, approaches based ML/DL appeal more and more interest of researchers. 
\IEEEpubidadjcol
In comparison with network architectures for single polarization case, those for multiple-polarization case always require a more delicate design~\cite{long2019geometrical, zhang2021ConvLSTM, pol2023li, wu2021polHRRP} although the building blocks may be similar.

On one hand, the traditional approaches are always limited by the mathematical/physical models they rely on. 
For example, the commonly employed scattering center model works well for targets where the distribution of scattering centers is fixed or pseudo-fixed. 
However, scattering centers are by nature a local description of scattering characteristics which is highly sensitive to the RLOS and polarization. 
In another word, the variation of the distribution of scattering centers with RLOS, polarization and etc., can be strong. 
How to figure out a remedy for the scattering center model to accommodate this variation is quite a hard task. 
On the other hand, the developed approaches based ML/DL can avoid the limitation arising from explicitly employing a mathematical/physical model as ML/DL inherently has the flexibility in establishing its own models/representations. 
Unfortunately, such a flexibility may make the associated approaches too expensive to be trained or even failing in finding a suitable model/representation to accomplish the required recognition task.

In this work, to alleviate the aforementioned challenges, a dual-polarization feature fusion network (DPFFN) is designed with a two-stage feature fusion strategy to integrate the global and local features of the dual-polarimetric HRRP sequences, where the understanding from the well-established scattering center models is exploited to the largest extent. 
The main contributions of this work are as follows:

\begin{enumerate}
    \item With the understanding of the scattering center model, a novel network called DPFFN is designed to extract and integrate multiple features from dual-polarimetric HRRP sequences. 

    \item A two-stage feature fusion strategy is proposed to realize two types of fusions where the global and local features in different polarized echos are exploited. 
    \item Based on the proposed feature fusion strategy, a new fusion loss function is developed to guide the network to generate comprehensive representations.

\end{enumerate} 

The rest of the paper is organized as follows. 
Section~\ref{sec:pol_hrrp} offers a brief introduction to polarimetric HRRP. 
Section~\ref{sec:proposed_network} describes the network architecture of the proposed DPFFN in detail, including the two-stage fusion strategy and the new fusion loss function. 
Section~\ref{sec:capability} presents the experimental results and provides a comparative analysis of the proposed network with existing ones. 
Section~\ref{sec:discussion} investigates the influence of various factors on the performance of the proposed DPFFN. 
Section~\ref{sec:conclusion} concludes the paper. 

\section{Polarimetric HRRP}\label{sec:pol_hrrp}
For RATR applications, it is suitable to view the scattering from complex targets as the superposition of fields generated by multiple distinct scattering centers~\cite{keller1962geometrical}. 
These centers correspond to specific regions of the target, such as corners, edges, and surface irregularities, which contribute differently to the overall scattering. 
Each scattering center is defined by a set of parameters determining its contribution to the scattered field, e.g., location, scattering amplitude, and angular scattering pattern~\cite{potter1995GTD}. 
The scattered fields and thus the radar echoes of a target can be modeled as the superposition of scattering centers, each contributing differently depending on the polarization. 
Taking the HH-polarization (horizontal transmit, horizontal receive) as an example, assuming $N$ scattering centers distributed along the RLOS (i.e., $x$-axis for simplicity), with each center located at a position $x_i$, the backscattered electric field, denoted as $E^{s}_{(HH)}(f)$, can be approximated as:
\begin{equation}
    E^{s}_{(HH)}(f) = \sum_{i=1}^{N} A^{i}_{(HH)} e^{-j2k\cdot x_i} = \sum_{i=1}^{N} A^{i}_{(HH)} e^{-j2\pi \frac{2f}{c} \cdot x_i}
\end{equation}
where $A^{i}_{(HH)}$ is the scattering amplitude of the $i$-th scattering center in the HH polarization and $k=2\pi f/c$ is the corresponding wave number for the working frequency $f$. 
The HH-polarized spatial range profile can be constructed by the inverse Fourier transforming as, 
\begin{equation}\label{eq:hrrp_ideal}
    \begin{aligned}
        E^{s}_{(HH)}(x) &= \mathcal{F}^{-1} [E^{s}_{(HH)}(f)]\\
    &=\int_{-\infty}^{\infty}[\sum_{i=1}^{N}A^{i}_{(HH)} \cdot e^{-j2\pi \frac{2f}{c} \cdot x_i}] \cdot e^{j2\pi \frac{2f}{c} \cdot x} d \frac{2f}{c}
    \end{aligned} 
\end{equation}
Here, it is assumed that the phase center of the scene is the geometrical center of the target. 

As the frequency bandwidth of a radar is limited, Eq.~\eqref{eq:hrrp_ideal} is always approximated as, 
\begin{equation}\label{eq:flh-f2}
    E^{s}_{(HH)}(x) = \frac{2B}{c} \sum_{i=1}^{N} A^{i}_{(HH)} \cdot e^{j2k_c(x-x_i)} \cdot sinc(\frac{2B}{c}(x-x_i))
\end{equation}
where $k_c=2\pi f_c/c$ is the wave number corresponding to the center frequency $f_c$, $B$ is the bandwidth of the radar system. 

Similar to the spatial range profile, the time-domain range profile is then constructed by applying the inverse Fourier transform as follows:
\begin{equation} \label{eq:hrrp_td}
    E^{s}_{(HH)}(t) = \mathcal{F}^{-1} [E^{s}_{(HH)}(f)] 
\end{equation}
A more general form of Eq.~\eqref{eq:flh-f2} is,
\begin{equation}
        \label{eq:polHRRP}
        E^{s}_{(HH)}(x) = \frac{2B}{c} \sum_{i=1}^{N} S_{(HH)}^i \cdot A^{i}_{(HH)} \cdot e^{j2k_c(x-x_i)} \cdot sinc(\frac{2B}{c}(x-x_i))
\end{equation}
where $S_{(HH)}^i$ gives the scattering intensity of the $i$-th scattering center for the HH polarization. 

For polarizations other than HH, i.e., HV, VV and VH, the range profiles can be similarly written as Eqs.~\eqref{eq:flh-f2} and~\eqref{eq:hrrp_td}. 
It is natural to define $S_{(HV)}^{i}$, $S_{(VV)}^{i}$, and $S_{(VH)}^{i}$ to describe the scattering intensities of the $i$-th scattering center for VH, VV, and HV polarizations, respectively. 
The full polarization information of the $i$-th scattering center can then be depicted by matrix $\mathbf{S}^{i}$, 
\begin{equation}
    \mathbf{S}^{i}=\left[\begin{array}{cc}
        S_{(HH)}^{i} & S_{(HV)}^{i} \\
        S_{(VH)}^{i} & S_{(VV)}^{i}
        \end{array}\right]
\end{equation}

\begin{figure}[htbp]
    \centering
    \includegraphics[width=0.35\textwidth]{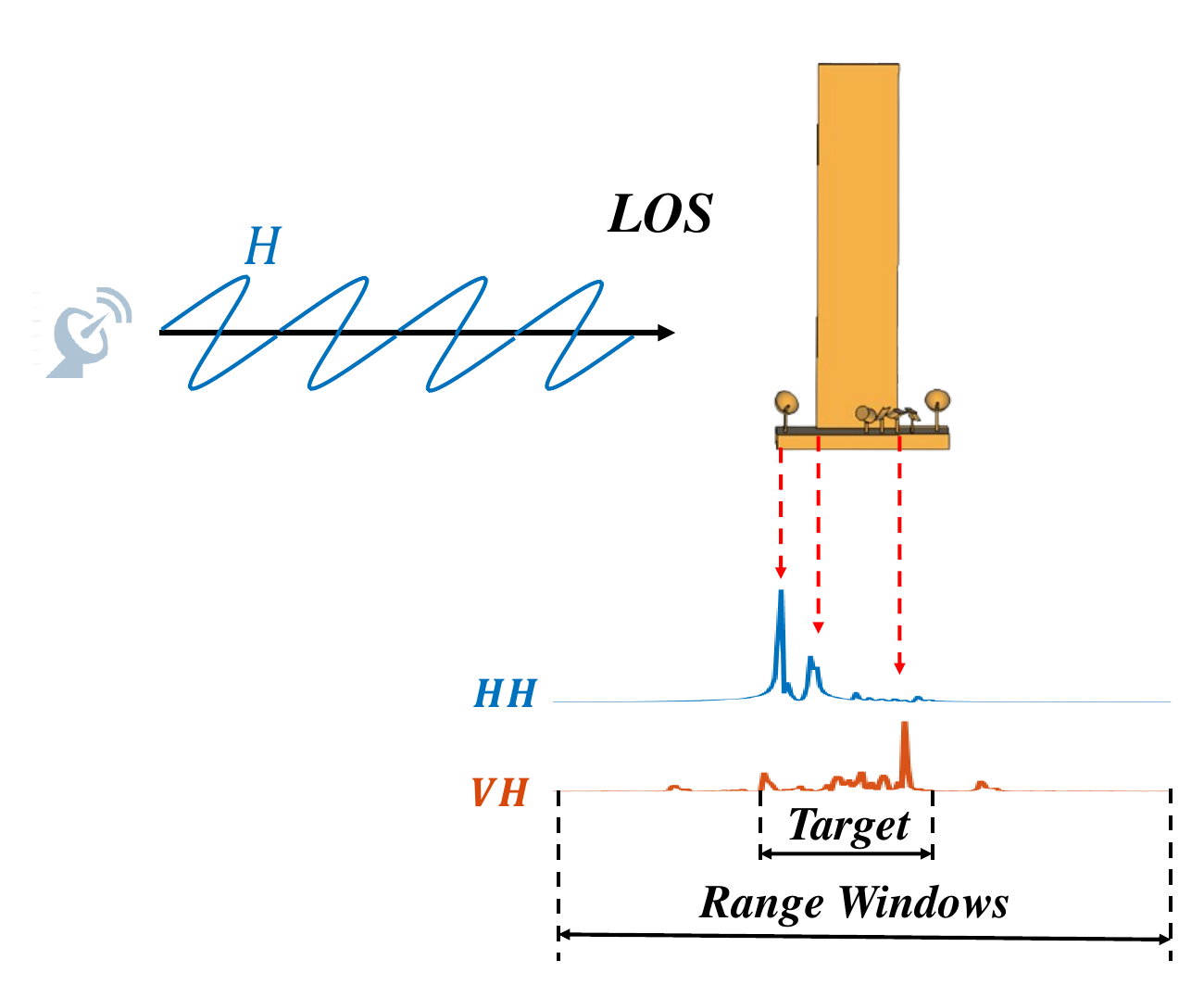}
    \caption{Polarimetric HRRPs of a satellite-like model}
    \label{fig:polHRRP}
\end{figure}

Polarimetric HRRPs contain echoes from different polarization channels~\cite{pol2015shengqi}.
An example of an HRRP for a satellite-like model is shown in Fig. \ref{fig:polHRRP}. 
As it is shown, the associated scattering centers contain rich scattering information from both the spatial and polarimetric viewpoints. 
In the spatial perspective, the HRRP accurately presents the distribution of the scattering centers along the RLOS. 
At the same time, it is obviously that scattering centers' distribution can vary greatly in different polarizations. 

As it is known, transmitting and receiving horizontally (H) and vertically (V) polarized waves simultaneously is quite challenging and expensive. 
For this reason, this work focuses on the case with a single H-polarized transmitter. 
To be more specific, HH- and VH-polarizations are investigated, although the proposed network can be extended to the fully polarized case. 

\section{Proposed Network}\label{sec:proposed_network}
This section begins with an overview of the DPFFN architecture, followed by a detailed explanation of its constituent modules. 

\subsection{Overview of the Network Architecture}
\begin{figure*}
    \centering
    \includegraphics[width=0.8\textwidth]{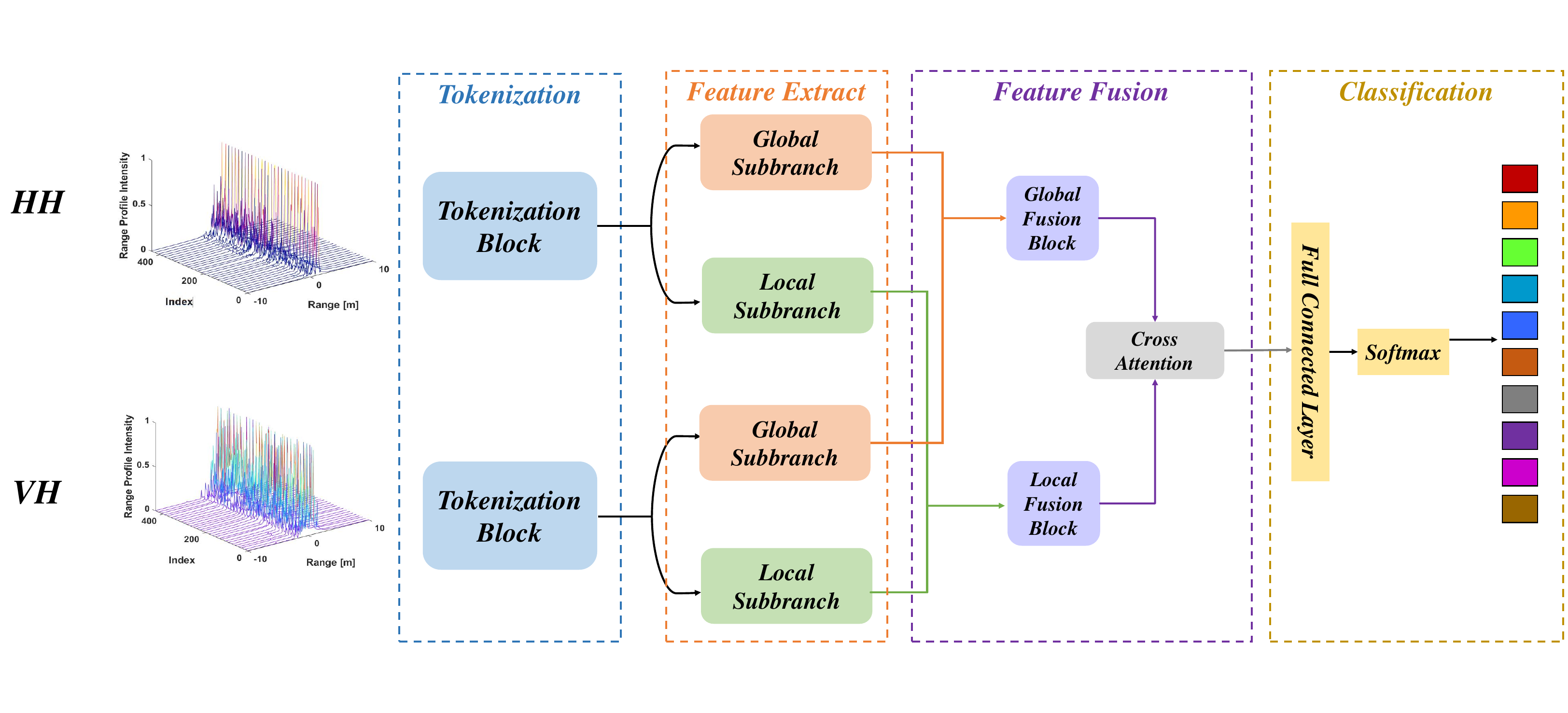}
    \caption{The overview of the proposed DPFFN}
    \label{fig:network}
\end{figure*}

Our DPFFN consists of four modules: the tokenization module, the feature extraction module, the feature fusion module, and the classification module, as shown in Fig.~\ref{fig:network}. 

The input dual-polarimetric HRRP sequences flows across the four modules in sequences.

The tokenization module transforms the continuous polarimetric HRRP sequences into tokenized representations, allowing for structured processing in the subsequent stages.

The feature extraction module extracts separately the global and local features by the corresponding global and local subbranches, which can then be exploited by the following feature fusion module.
As we know, scattering centers themselves are a local description of a radar echo.
an HRRP, which consists of a set of scattering centers, can be viewed as a global representation of the whole scatterer/target of interest.
At the same time, the distribution and structure of an HRRP always varies slowly with the observation direction. 
The direct consequence is that HRRPs with respect to adjacent views reflect invariance characteristics of the scatterer. 
This understanding explains our design of the feature extraction module. 
To the best of the authors' knowledge, such a design is not available in the literature. 

As its name indicated, the feature fusion module fuses the obtained features generated by its preceding module. 
Different from other polarimetric fusion networks (network modules)~\cite{zhang2021ConvLSTM,pol2023li}, the proposed module realizes two types of fusions by a two-stage procedure. 
The first stage employs two branches, handling the global and local features separately. 
The second stage utilizes the cross-attention to fuse global and local features. 

Finally, the classification module accepts the fused features and conducts the prediction task. 

\subsection{Tokenization Module}
\begin{figure}[!htbp]
    \centering
    \includegraphics[width=0.48\textwidth]{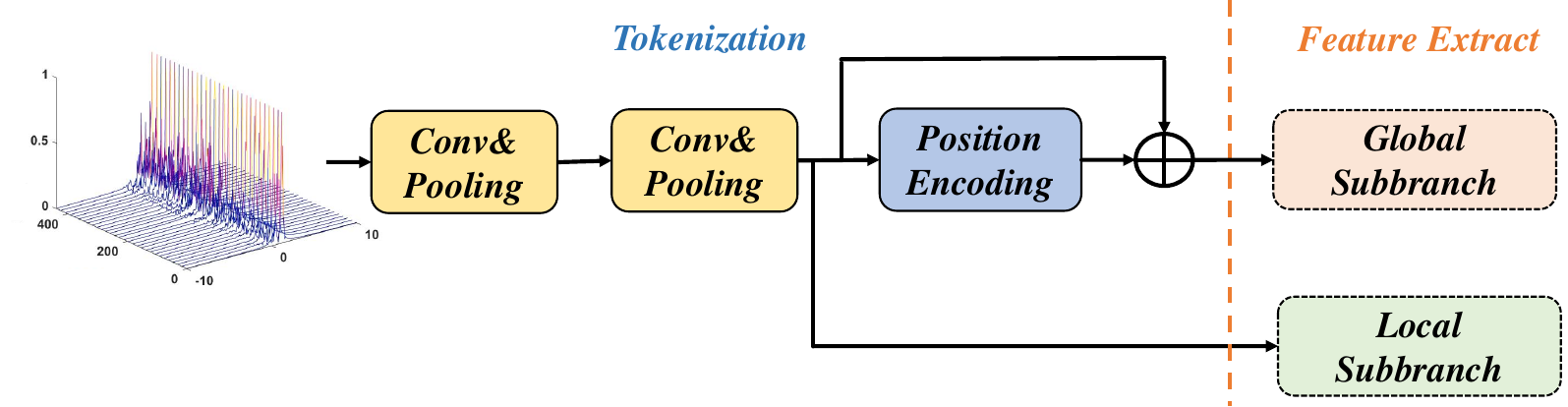}
    \caption{The Tokenization Module. Polarimetric HRRP sequences are processed through two CNN layers to extract foundational features. These features are passed to the position encoding layer for token generation used by the global subbranch in feature exaction module, while the local subbranch directly receives the CNN output. }
    \label{fig:token}
\end{figure}

The tokenization module transforms the input sequences into the desired tokens. 
Since two polarizations to be handled, the tokenization module has two identical branches, as shown in Fig.~\ref{fig:token}. 
Here, each branch consists of two convolutional neural networks (CNN) layers and a position encoding layer, similar to that in~\cite{spaceHRRP2023zhang}. 
As will be discussed in Section~\ref{sec:feature_extraction_module}, the global subbranch in the feature extraction module requires the positional information except for the feature maps themselves. 
Simultaneously, the local branch requires only the feature maps alone.
To accommodate this difference, the tokenization module will send the obtained feature maps directly to the local branch of the feature extraction module by skipping the positional encoding layer. 
Meanwhile, the positional encoding layer will be activated to include the positional information into the feature maps required by the global subbranch of the feature extraction module.

As discussed in Section~\ref{sec:intro}, the spatial pattern of an HRRP is heavily dependent on the target's scattering centers, which are the basis to realize RATR. 
In this work, CNN layers are used to extract the local spatial features of the HRRP before generating the positional tokens, which are required by the following transformer encoder in the feature exaction module. 
The underlying rationale for choosing CNN is that the scattering centers are inherently a local description of the scattering characteristic. 
With the CNN layers, the obtained tokens can focus more on the local spatial features. 
Along with the following transformer encoder which excels at capturing the global features, such a design makes the proposed network perform well in extracting both local and global information in HRRPs.

\subsection{Feature Extraction Module}\label{sec:feature_extraction_module}
Similarly to the aforementioned tokenization module, the feature extraction module has two branches, each for a single polarization. 
Each branch here can be divided into two subbranches: one for global features and the other for local features. 
As they are named, the former is responsible for capturing the global characteristics, while the latter for fine-grained local details from each sequence.

\subsubsection{Global Subbranch}\label{sec:global}
Inspired by the Vision Transformer (ViT)~\cite{dosovitskiy2020image}, the proposed network employs the transformer encoder as the building block. 
The distribution and structure of scattering centers always vary slowly with the observation directions. 
A direct consequence is that, from HRRPs within adjacent views, the invariance characteristics can be obtained associated with the scatterer. 
The proposed network make use of the transformer encoder to extract these information. 

\begin{figure}
    \includegraphics[width=0.48\textwidth]{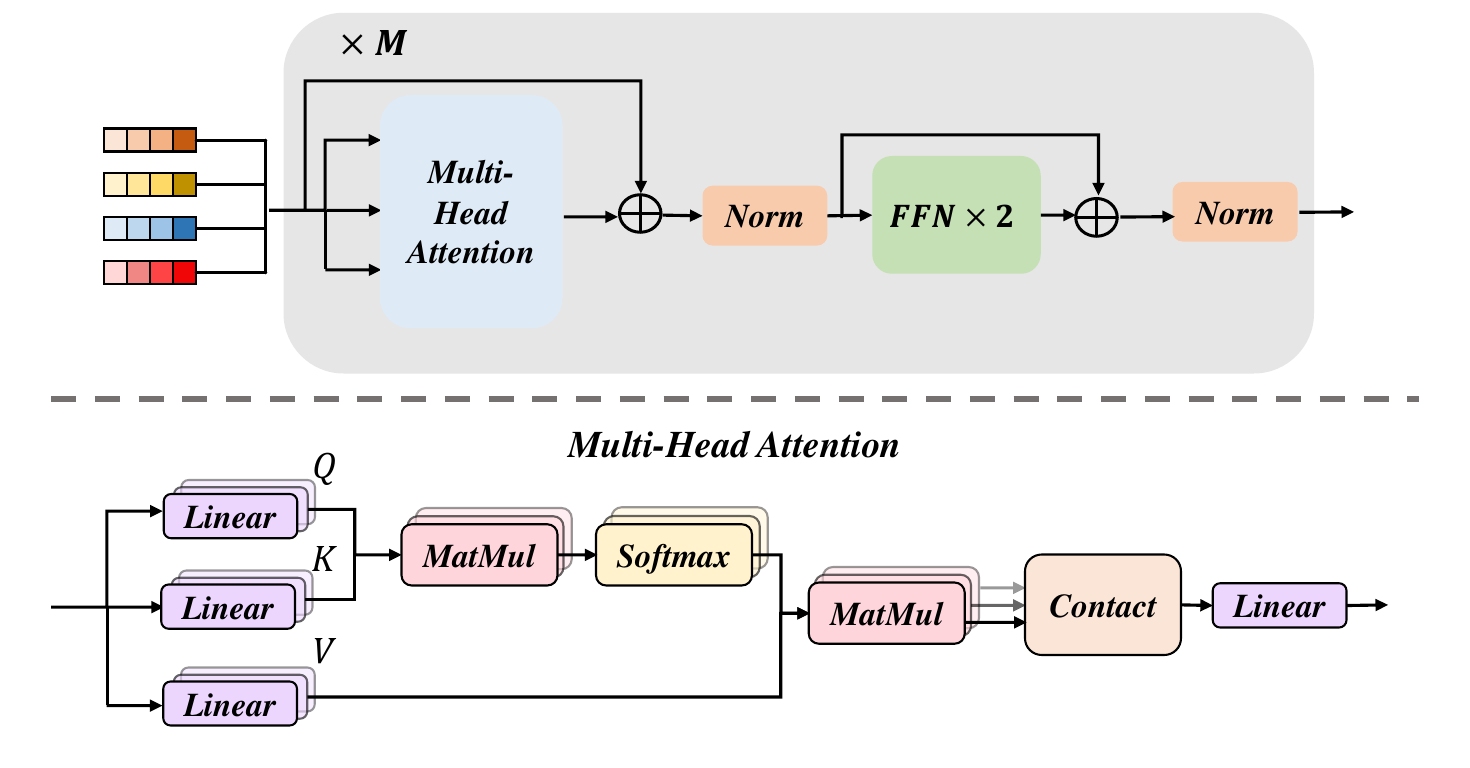}
    \caption{The global subbranch. This subbranch consists of $M$ transformer encoder layers, which are specifically designed to extract global features from the HRRP sequences. }
    \label{fig:global}
\end{figure}

As illustrated in Fig.~\ref{fig:global}, this branch consists of $M$ specifically designed transformer encoders. 
The rule to choose $M$ is the trade-off between the computational efficiency and the feature extraction capacity, as would be investigated in Section~\ref{sec:hyperparam}.

\subsubsection{Local Subbranch}

The self-attention mechanism excels at capturing global features but is inherently limited in extracting detailed local features~\cite{peng2021conformer}. 
To overcome this limitation, the DPFFN designs a so-called local subbranch to capture the detailed local features within an HRRP. 

\begin{figure}[t]
    \centering
    \includegraphics[width=0.35\textwidth]{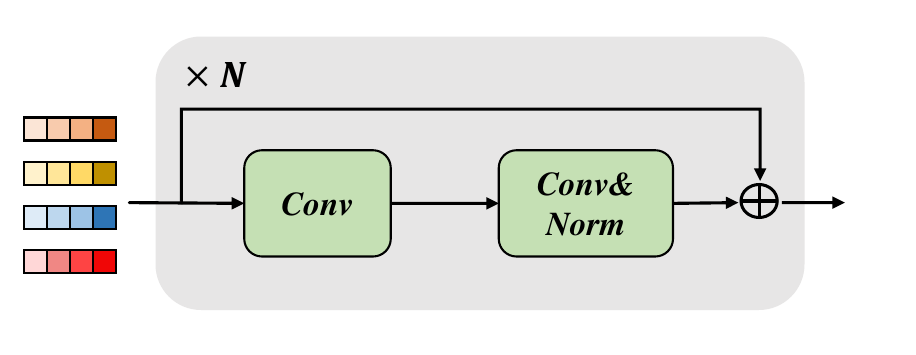}
    \caption{The local subbranch. This subbranch consists of $N$ residual CNN blocks, which are specifically designed to extract local features from the HRRP sequences. }
    \label{fig:local}
\end{figure}

Inspired by ResNet~\cite{he2016resnet}, the local module subbranch uses stacked CNN blocks with residual connections, as depicted in Fig.~\ref{fig:local}. 
This subbranch consists of $N$ specifically designed CNN layers. 
Similar to the choice of $M$, $N$ is selected to balance the computational efficiency and the feature extraction capacity. 

With the combination of the global and local modules, both global and local features of the HRRP sequences can be successfully extracted. 
It is ready now to fuse the features from the dual polarizations. 

\subsection{Feature Fusion Module}
Essentially, the feature fusion module joins the features generated by the preceding feature extraction modules. 
Different from conventional feature fusion strategies where features from different polarizations~\cite{zhang2021ConvLSTM,pol2023li} are simply concatenated, two types of fusions are conducted here by a two-stage process, as shown in Fig.~\ref{fig:network}. 
In the first stage, global and local features with respect to the two polarizations are separately fused. 
The obtained two set of features are then fused by a cross attention. 
In addition, both types of fusions in this work are conducted by an adaptive and flexible manner, where a dynamical adjustment of the parameters associated with the fusion operations is enabled according to the extracted characteristics of the target. 
As a result, all available features are made good use to the largest extent. 

\begin{figure}
    \centering
    \includegraphics[width=0.48\textwidth]{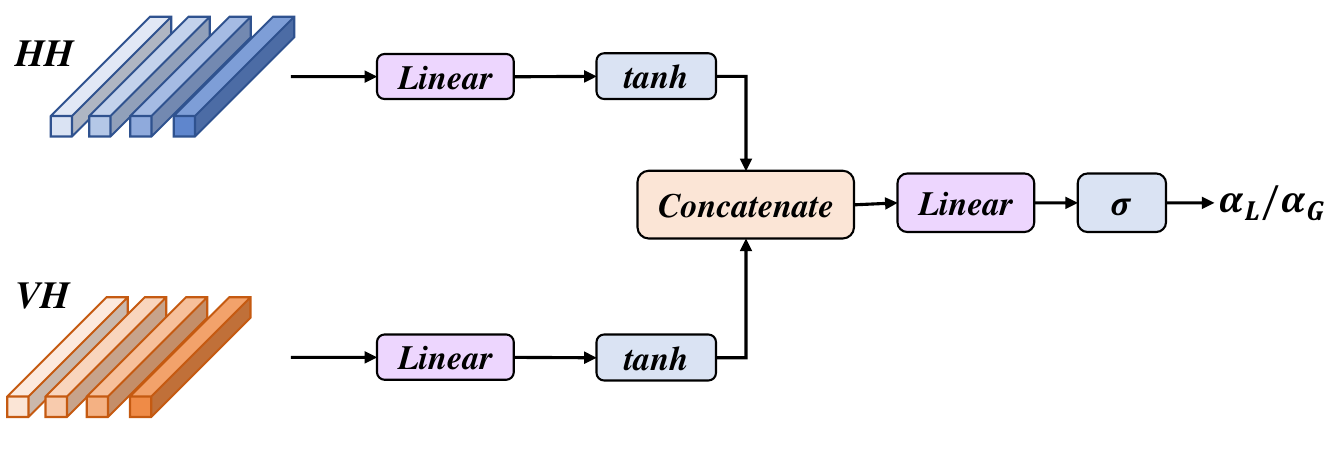}
    \caption{The First Stage Feature Fusion Block. This block uses a GMU to integrate global and local features from HH and VH polarizations, respectively. }
    \label{fig:gated}
\end{figure}

\subsubsection{First Stage}
The first stage consists of two blocks, the global fusion block and the local fusion block. 
Since the two blocks sharing a same structure, only one of them is presented in Fig.~\ref{fig:gated}, and the following discussion will take one of them, namely, the global fusion block, as an example to explain them. 
The adaptive and flexible fusion in the first stage is realized from two aspects.

From one aspect, the block utilizes gated multimodal units (GMU)\cite{arevalo2017gated} to integrate the global features from HH and VH polarizations, which employs a gating mechanism to dynamically control the integration process. 
By assigning adaptive weights to different data sources, the GMU effectively balances the contributions from these data during fusion. 

From the other aspect, two fusion weights $\alpha_{G}$ and $\alpha_{L}$ are employed for the global and local features.
As shown in Fig.~\ref{fig:gated}, the $\tanh$ activation function is applied to the linear transformations of the input features, generating the associated latent representations. 
The obtained representations are then concatenated and subjected to another linear transformation followed by a sigmoid activation to compute the fusion weight $\alpha_{G}$. 
The final fused output $F_{G}$ is a weighted sum of the original global features with a automatically adjusted parameter $\alpha_{G}$. 
The output of the global fusion block can be written as,
\begin{equation}
    F_{G} = \alpha_{G} \cdot F_{(HH)}^{G} + (1-\alpha_{G}) \cdot F_{(VH)}^{G}
\end{equation}
where $F_{(HH)}^{G}$ and $F_{(VH)}^{G}$ represent the global features of the HH and VH polarizations, respectively, and $\alpha_{G}$ signifies the self-adaptive fusion weight for the global features of the HH and VH polarizations. 

As far as the local fusion block is concerned, the output is denoted by $F_{L}$. 
The output of the local fusion block can be written as,
\begin{equation}
    F_{L} = \alpha_{L} \cdot F_{(HH)}^{L} + (1-\alpha_{L}) \cdot F_{(VH)}^{L}
\end{equation}

\subsubsection{Second Stage}
\begin{figure}[t]
    \centering
    \includegraphics[width=0.48\textwidth]{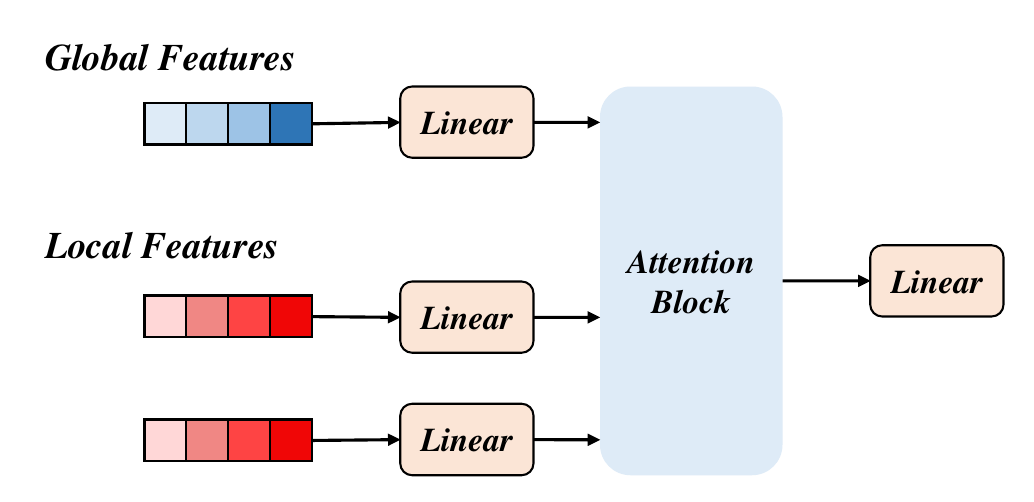}
    \caption{The Second Stage Feature Fusion Block. This block uses a cross-attention mechanism to integrate global and local features. }
    \label{fig:cross}
\end{figure}

The second stage employs a cross-attention mechanism to capture interactions between the fused global $F_{G}$ and local features $F_{L}$, as shown in Fig.~\ref{fig:cross}. 

The block in this stage utilizes a cross-attention mechanism \cite{crossvit} to handle the global and local features, which enables the model to focus on relevant regions of both feature sets, dynamically capturing intricate relationships between global and local representations. 
Such a mechanism is particularly valuable in polarimetric HRRP-based applications, where the interplay between global and local features is critical for distinguishing closely related target classes.

In summary, two types of fusions are realized by a two-stage process in this work. 
Both stages are conducted by an adaptive and flexible manner.

\subsection{Classification Module}
The classification module is responsible for generating the classification results. 
It comprises of a fully connected layer followed by a softmax layer. 
The fully connected layer reduces the dimensionality of the fused features, while the softmax layer produces the final classification predictions. 

\subsection{Loss Function}
In this work, the loss, denoted by $\mathcal{L}_{total}$ here, is defined as:

\begin{equation} 
    \mathcal{L}_{total} = \mathcal{L}_{cross} + \lambda \cdot \mathcal{L}_{fusion} \label{eq:loss_function},
\end{equation}
with 
\begin{subequations} 
    \begin{align}
    \mathcal{L}_{cross} &= -\sum_{i=1}^{N} y_{i} \cdot \log(\hat{y}_{i}) \label{eq:loss_function_cross},\\
    \mathcal{L}_{fusion} &= \frac{|R(F_{(HH)}^{L},F_{(VH)}^{L})|}{R(F_{(HH)}^{G},F_{(VH)}^{G})+\epsilon}. \label{eq:loss_function_fusion}
    \end{align}
\end{subequations}
In Eq.~\eqref{eq:loss_function}, $\mathcal{L}_{cross}$ represents the cross-entropy loss, $\mathcal{L}_{fusion}$ denotes the fusion loss, and $\lambda$ is a hyperparameter that balances these two terms. 
It is set to 2 in this work according to our numerical experience. 
In Eq.~\eqref{eq:loss_function_cross}, $y_{i}$ and $\hat{y}_{i}$ are the ground-truth and predicted labels, respectively. 
In Eq.~\eqref{eq:loss_function_fusion}, $R$ is the correlation coefficient. 
The parameter $\epsilon$ is set to 1.01 to ensure $\mathcal{L}_{fusion}$ being a positive number within a stable range. 

In Eq.~\eqref{eq:loss_function}, the term $\mathcal{L}_{fusion}$ distinguishes our proposed loss function from the ones employed in existing networks. 
The term $\mathcal{L}_{cross}$ is commonly used in existing networks, quantifying the discrepancy between predicted and ground-truth labels. 
In contrast, the term $\mathcal{L}_{fusion}$ is uniquely designed in this work to explicitly exploit the distinct characteristics of dual polarimetric echoes. 
While the global features represent fundamental attributes such as shape, size, and electromagnetic properties—attributes that are consistent across polarizations, the local features reflect fine details of the scatterer which are always polarization sensitive. 
As a result, global features from HH and VH polarizations tend to be positively correlated, while local features can exhibit significant differences. 
By aligning the global features and promoting differentiation in local features, $\mathcal{L}_{fusion}$ enables the model to learn both the shared and unique characteristics of different polarizations associated with the target.

\section{Capability}\label{sec:capability}
This section shows the capability of the proposed DPFFN by a comparative study against serval commonly used networks. 

\subsection{Dataset}

\begin{figure*}[htbp]
    \subfloat[\centering class 0]{\includegraphics[scale=0.32, trim=0 0 0 0,clip]{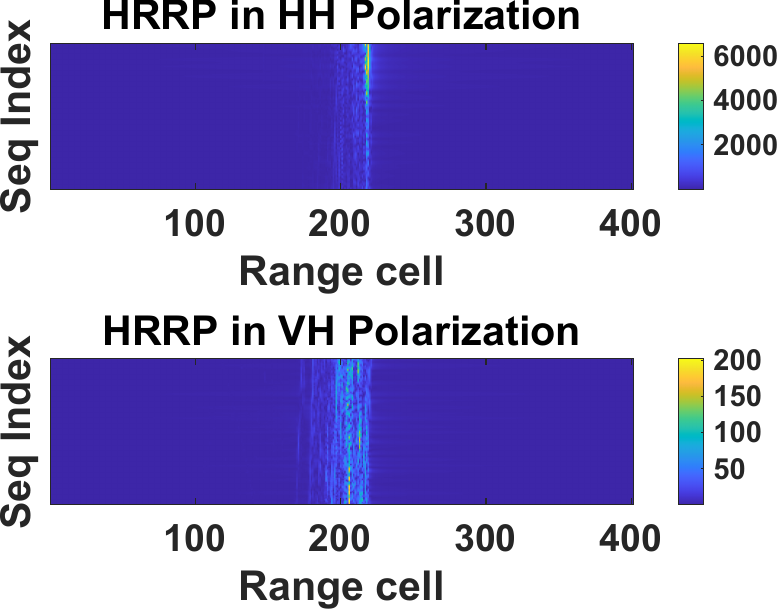}}
    \hspace{0.3cm}
    \subfloat[\centering class 1]{\includegraphics[scale=0.32, trim=0 0 0 0,clip]{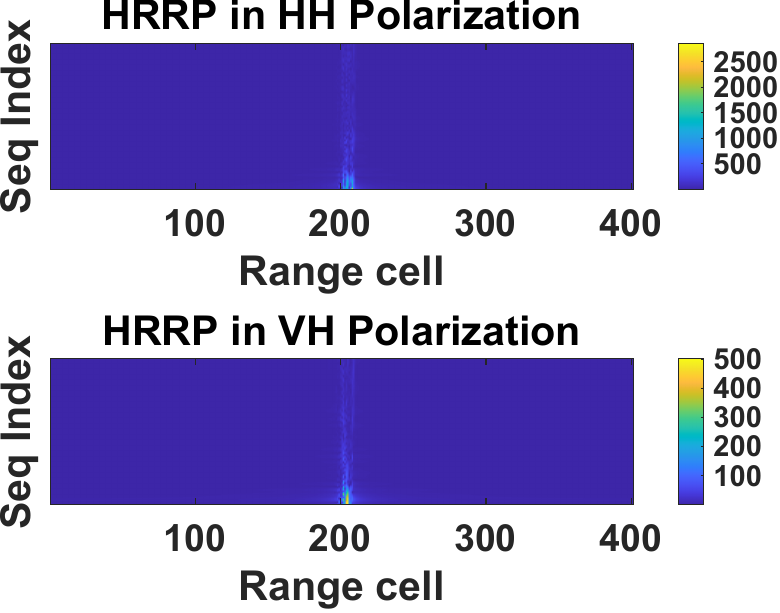}}
    \hspace{0.3cm}
    \subfloat[\centering class 2]{\includegraphics[scale=0.32, trim=0 0 0 0,clip]{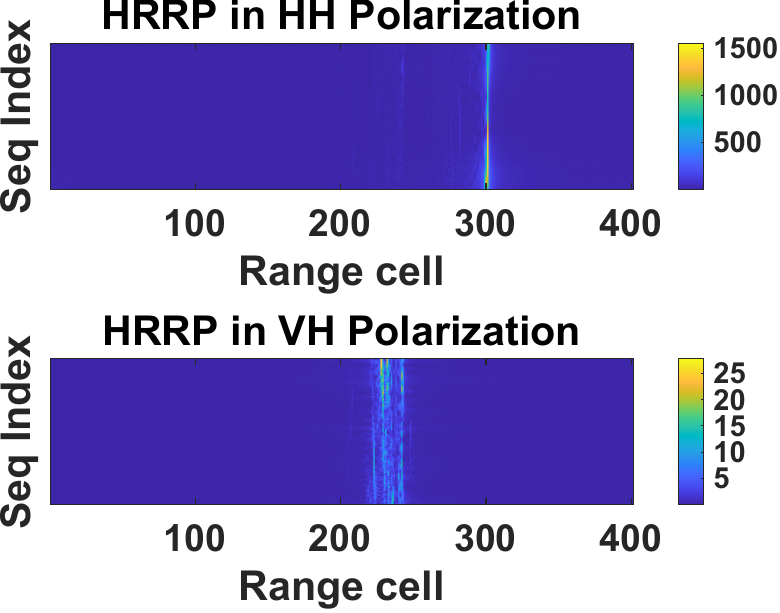}}
    \hspace{0.3cm}
    \subfloat[\centering class 3]{\includegraphics[scale=0.32, trim=0 0 0 0,clip]{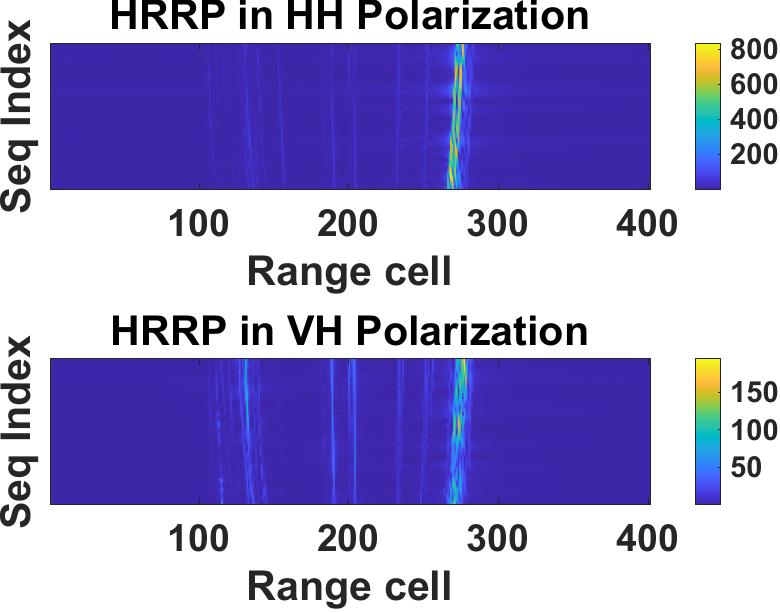}}
    
    \subfloat[\centering class 4]{\includegraphics[scale=0.32, trim=0 0 0 0,clip]{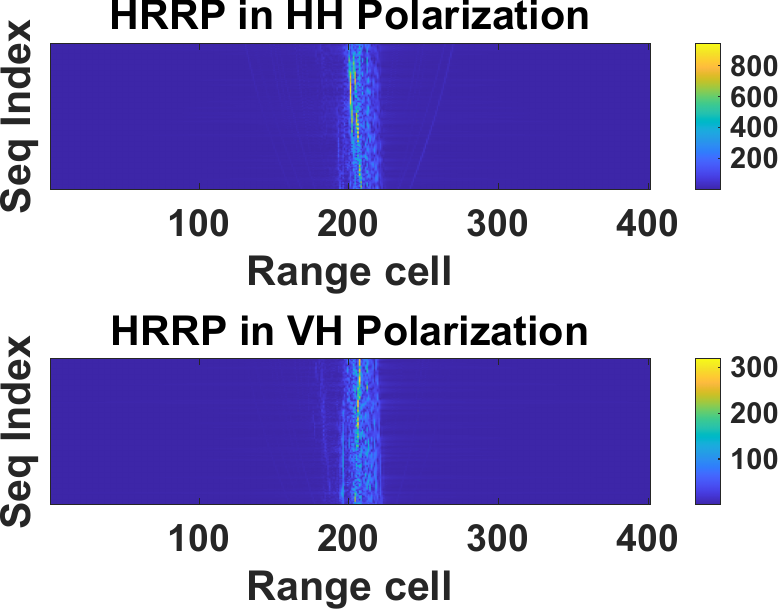}}
    \hspace{0.3cm}
    \subfloat[\centering class 5]{\includegraphics[scale=0.32, trim=0 0 0 0,clip]{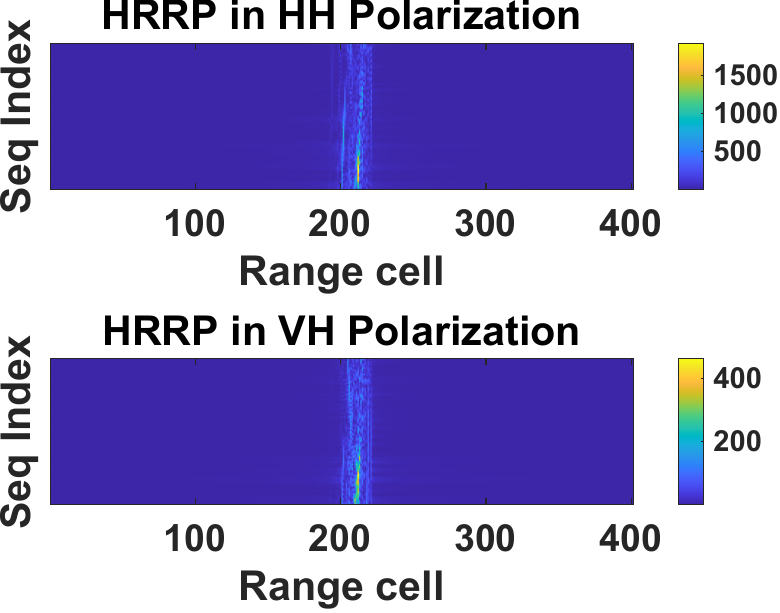}}
    \hspace{0.3cm}
    \subfloat[\centering class 6]{\includegraphics[scale=0.32, trim=0 0 0 0,clip]{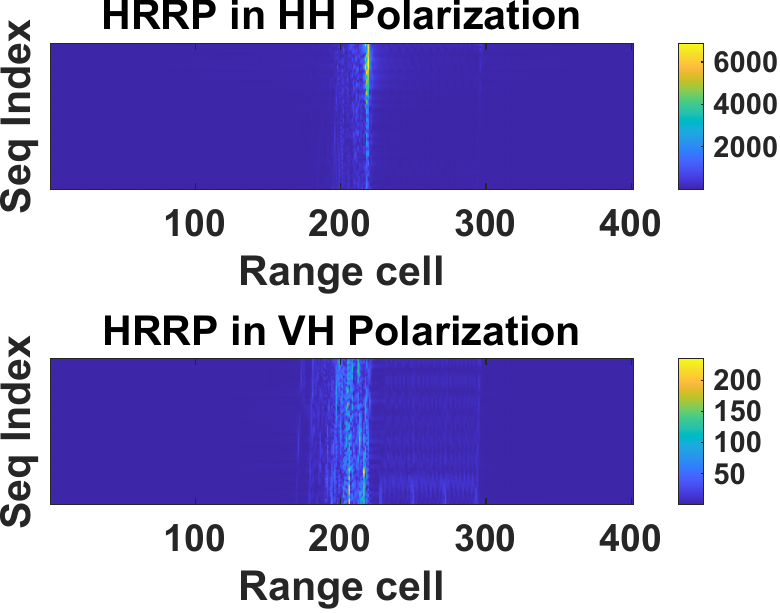}}
    \hspace{0.3cm}
    \subfloat[\centering class 7]{\includegraphics[scale=0.32, trim=0 0 0 0,clip]{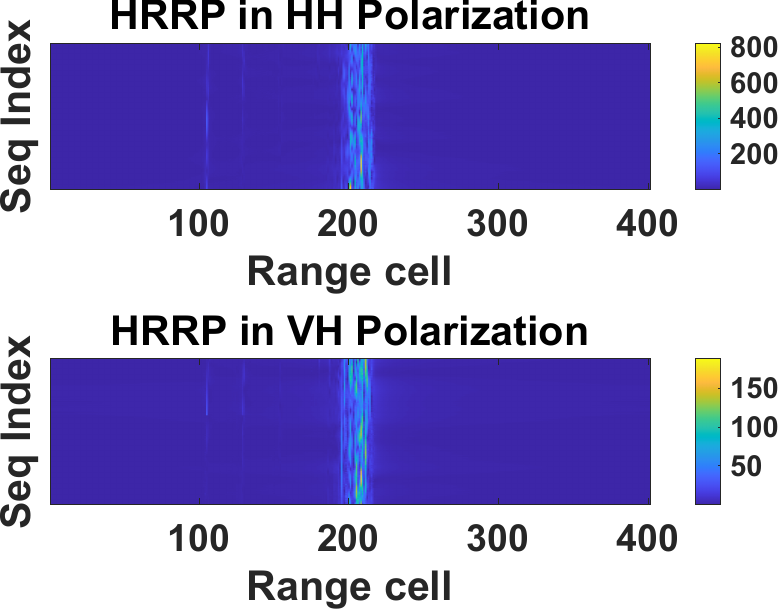}}
    
    \subfloat[\centering class 8]{\includegraphics[scale=0.32, trim=0 0 0 0,clip]{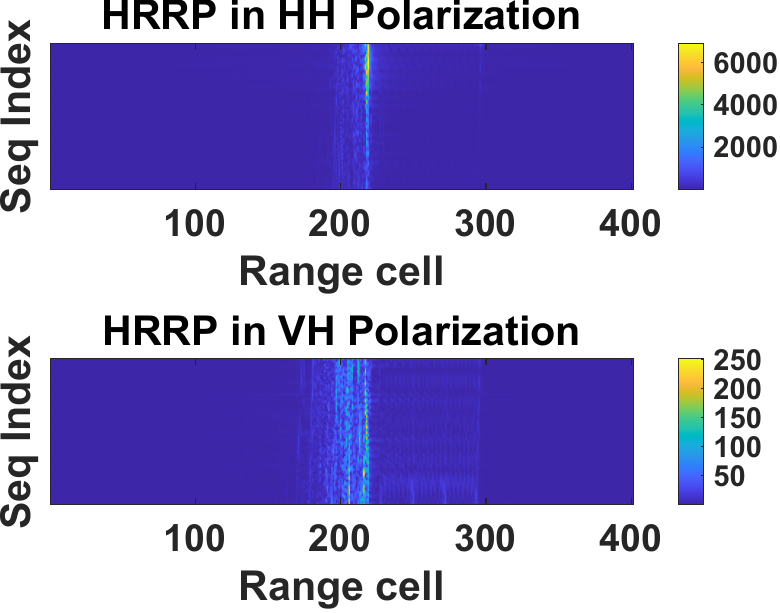}}
    \hspace{0.3cm}
    \subfloat[\centering class 9]{\includegraphics[scale=0.32, trim=0 0 0 0,clip]{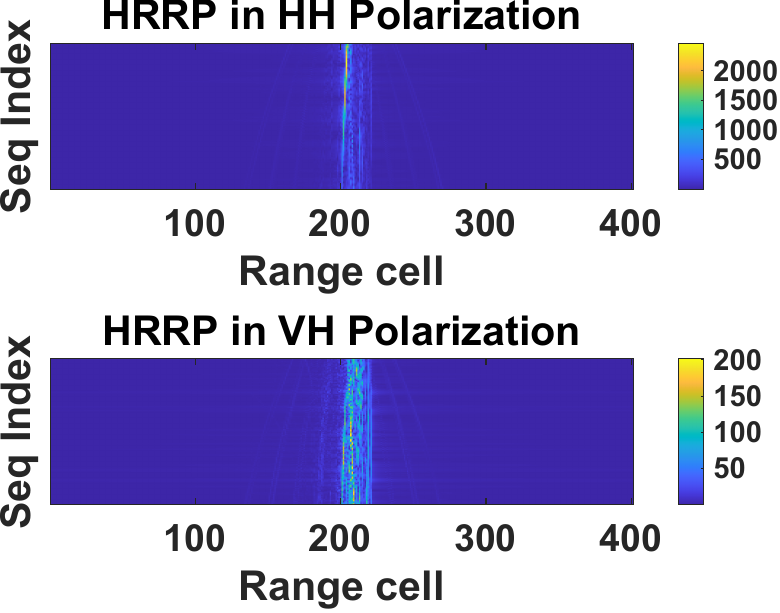}}

    \caption{The visualization of the dataset, from (a) to (j) are the HRRPs of class 0 to class 9 satellite models.}
    \label{fig:dataset}
\end{figure*}

The dataset consists of simulated polarimetric HRRP sequences of ten satellite models, each having 10 postures. 
For a satellite model, 25 HRRP sequences are generated for each posture, where an HRRP sequence has 512 HRRPs. 
As a result, the dataset contains 2500 HRRP sequences. 
The parameters of the radar settings are presented in Table~\ref{tab:radar}.
Figure~\ref{fig:dataset} shows the randomly selected HRRPs of the ten models. 
In following experiments, HRRP sequences associated randomly chosen 8 postures of each model are taken as training set. 

\begin{table}[htbp]
    \centering
    \caption{The parameters of the incident wave in our dataset.}
    \label{tab:radar}
    \begin{tabular}{c|c}
        \hline
        \textbf{Parameter} & \textbf{Value} \\
        \hline
        Center Frequency & 9 GHz \\
        Bandwidth & 2 GHz \\
        Step Frequency & 5 MHz \\
        \hline
    \end{tabular}
\end{table}

\subsection{Setup}
All experiments are conducted on a server with an AMD EPYC 9654 CPU and an NVIDIA GeForce RTX 4090 GPU where Ubuntu 22.04, Python 3.12 and PyTorch 2.3.0 are equipped. 
In the feature extraction module of the proposed DPFFN, the global subbranch has 10 transformer encoder layers, each owning 10 attention heads, meanwhile, the corresponding local subbranch owns 10 concatenated residual blocks. 
The initial learning rate is set to 0.001, which is reduced by a factor of 0.8 every 50 epochs to improve the convergence of training. 
The training batch size is set to 32. 
The maximum number of epochs is 300 during training, with a periodic validation to monitor performance metrics. 

As the primary metric to evaluate the classification performance, the accuracy rate is defined as the ratio of correctly classified samples to the total number of samples. 
To evaluate the performance of the investigated networks, the capacity and computational efficiency are investigated against the model parameters (Params) and the floating-point operations per second (FLOPs). 

\subsection{Results}\label{sec:capability_results}

The performance of the DPFFN is compared with several representative networks, including that based on convolutional neural network (CNN) \cite{Resnet2019Guo, CNN2019Guo}, recurrent neural network (RNN)~\cite{RNN2019Xu, RNN2021Chen, RNN2019Liu}, long short-term memory network (LSTM) \cite{zhang2021ConvLSTM}, and on Transformer~\cite{vaswani2017attention,spaceHRRP2023zhang,wang2023high}. 
As the above networks were all originally developed to handle the case with a single polarization, only HH-polarization is employed here for the comparative study.
It is interesting to point out that the Transformer employed here has the same architecture as the global subbranch of our DPFFN (without the local subbranch). 
Of course, the polarization fusion operation and fusion loss term are both missing in this Transformer since only one polarization is employed. 
Due to this similarity, Transformer will be used as the baseline network for the ablation investigation in the following Section~\ref{sec:ablation_studies}.

\begin{table*}[htbp]
    \centering
    \caption{Comparision of the investigated networks.}
    \label{tab:result}
    \begin{tabular}{c|c|c|c|c}
        \hline
        & \textbf{Description} & \textbf{Accuracy} & \textbf{Params} ($10^6$) & \textbf{FLOPs} ($10^6$)\\
        \hline
        CNN & ResNet-18 & 73.8\% & 4.03 & 133.72\\
        RNN & Stacked RNNs & 71.8\% & 23.66 & 13094.04\\
        LSTM &Stacked Bi-LSTMs & 83.2\% & 4.23 & 480.51\\
        Transformer& Similar to Global Subbranch of Our DPFFN & 88.4\% & 8.32 & 2421.43\\
        \cyanbf{Ours}& Our DPFFN & \textbf{94.8\%} & 16.69 & 5420.10\\
        \hline
    \end{tabular}
\end{table*}

Table~\ref{tab:result} details the accuracy, parameters count, and FLOPs for the investigated networks. 
Here, the dataset is noise-free. 
Obviously, the proposed DPFFN has the highest accuracy and superior performance. 
In particular, 
\begin{itemize} 
    \item The proposed DPFFN achieved the highest accuracy of 94.8\%. As will be shown in Section~\ref{sec:ablation_studies}, the nice performance can be attributed to the proposed two-stage feature fusion strategy, which successfully two types of fusions by exploiting global and local features from multiple polarizations. 
    \item The number of parameters in the proposed DPFFN is 16.69 millions and the corresponding FLOPs reaches to 5.42 billions. 
    They are significantly larger than other networks except for the RNN network, due to its two-branch architecture for dual polarizations. 
\end{itemize}

\begin{figure*}[htbp]
    \subfloat[\centering CNN ]{\includegraphics[scale=0.44, trim=0 0 120 0,clip]{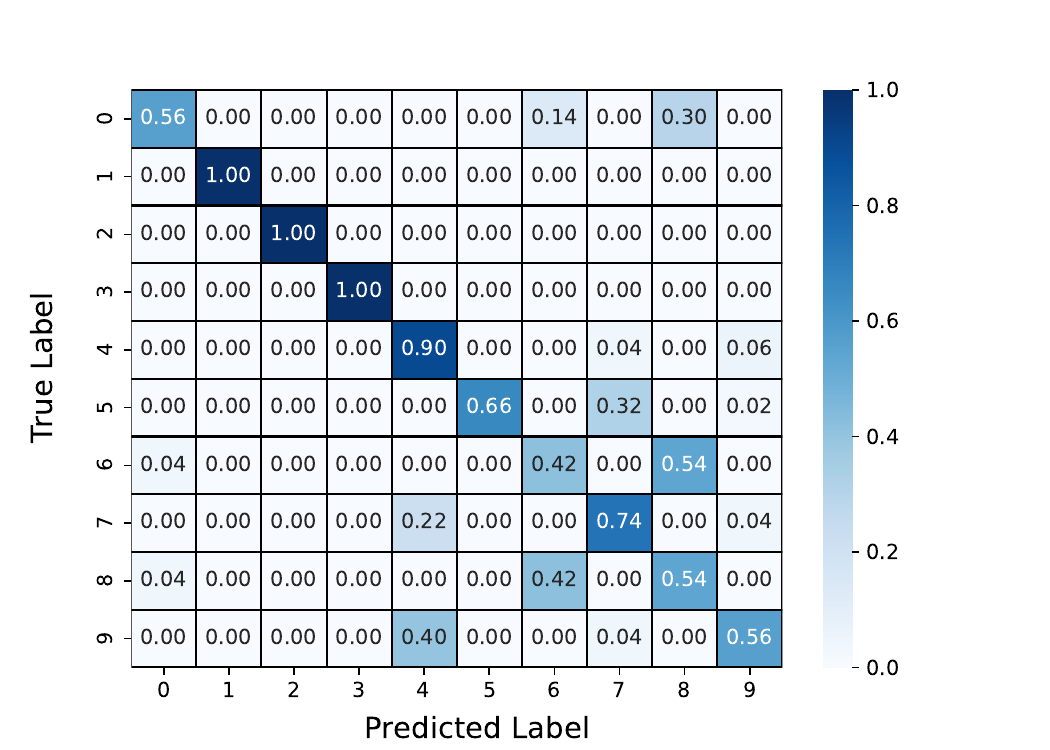}}
    \subfloat[\centering RNN ]{\includegraphics[scale=0.44, trim=0 0 120 0,clip]{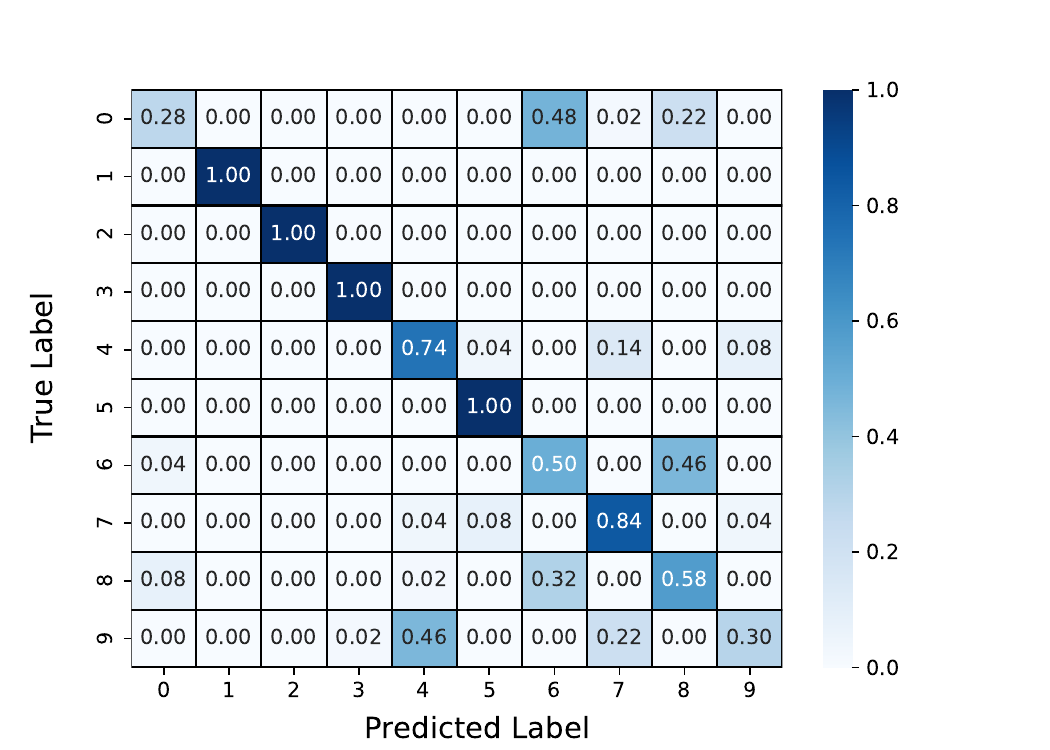}}
    \subfloat[\centering LSTM ]{\includegraphics[scale=0.44, trim=0 0 120 0,clip]{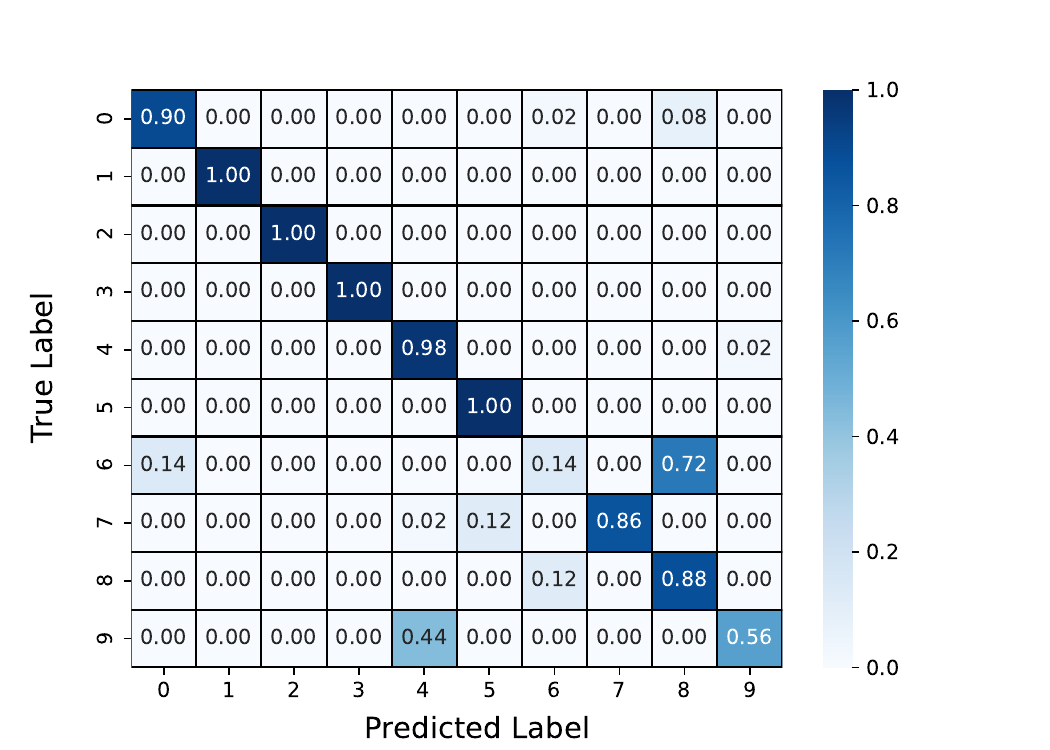}}
    \\
    \subfloat[\centering Transformer]{\includegraphics[scale=0.44, trim=0 0 120 0,clip]{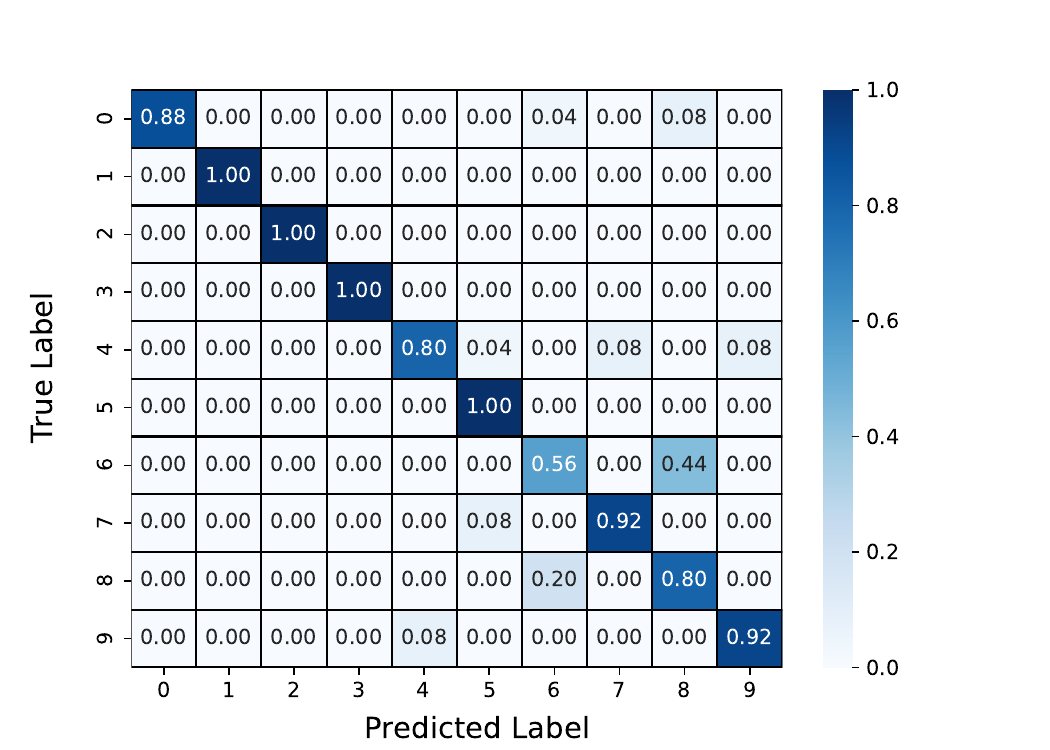}}
    \subfloat[\centering \textbf{Ours}]{ \includegraphics[scale=0.44, trim=0 0 120 0,clip]{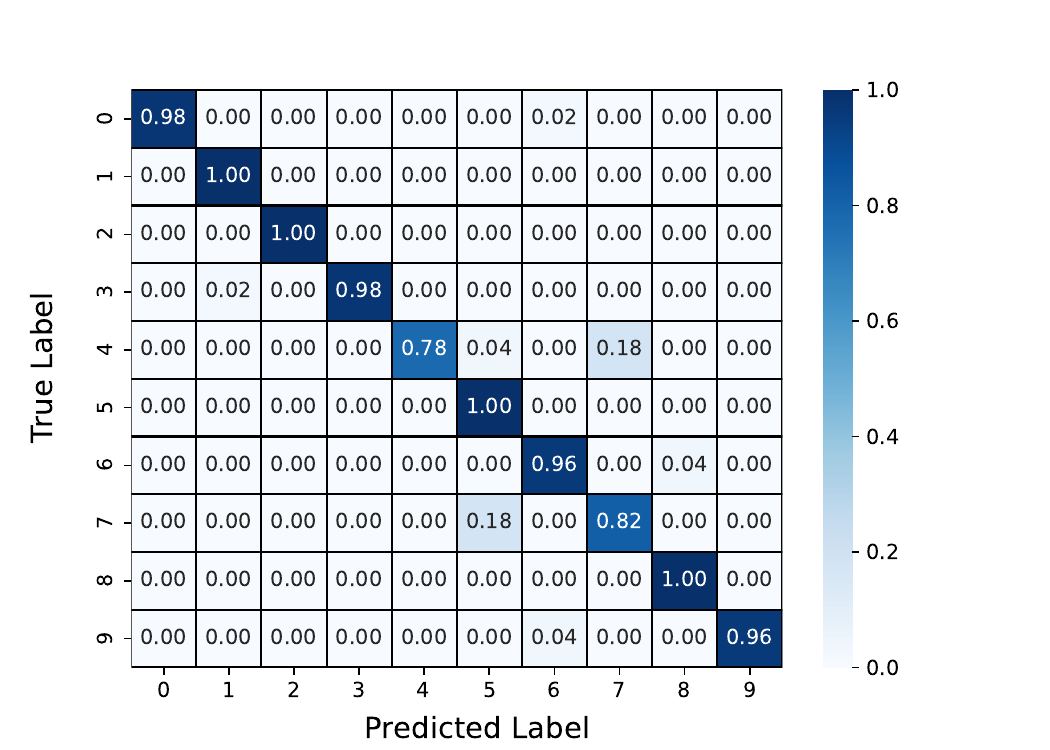}}
    \caption{The confusion matrices of the investigated networks.
    }
    \label{fig:confusion}
\end{figure*}

To provide a more detailed comparison, confusion matrices for all networks are presented in Fig.~\ref{fig:confusion}.
The results show that, 
\begin{itemize} 
    \item The CNN network achieves a good accuracy for satellite models 1, 2, 3, and 4. 
    \item The RNN network performs well for satellite models 1, 2, 3, and 5. 
    \item The LSTM network demonstrates a good accuracy for satellite models 0, 1, 2, 3, 4, 5, 7, and 8, but shows poor performance for satellite models 6 and 9. 
    \item Transformer network excels in accuracy for satellite models 0, 1, 2, 3, 4, 5, 7, 8, and 9, but performs poorly for satellite model 6. 
    \item The DPFFN network achieves good accuracy for nearly all satellite models, with a slight degradation of the accuracy for models 4 and 7. 
\end{itemize}

Our DPFFN shows superior ability to identify these targets at a very high accuracy, making it a promising candidate for future research in RATR tasks. 

\begin{figure*}[htbp]
    \centering
    \subfloat[\centering CNN ]{\includegraphics[scale=0.3, trim=0 0 120 0,clip]{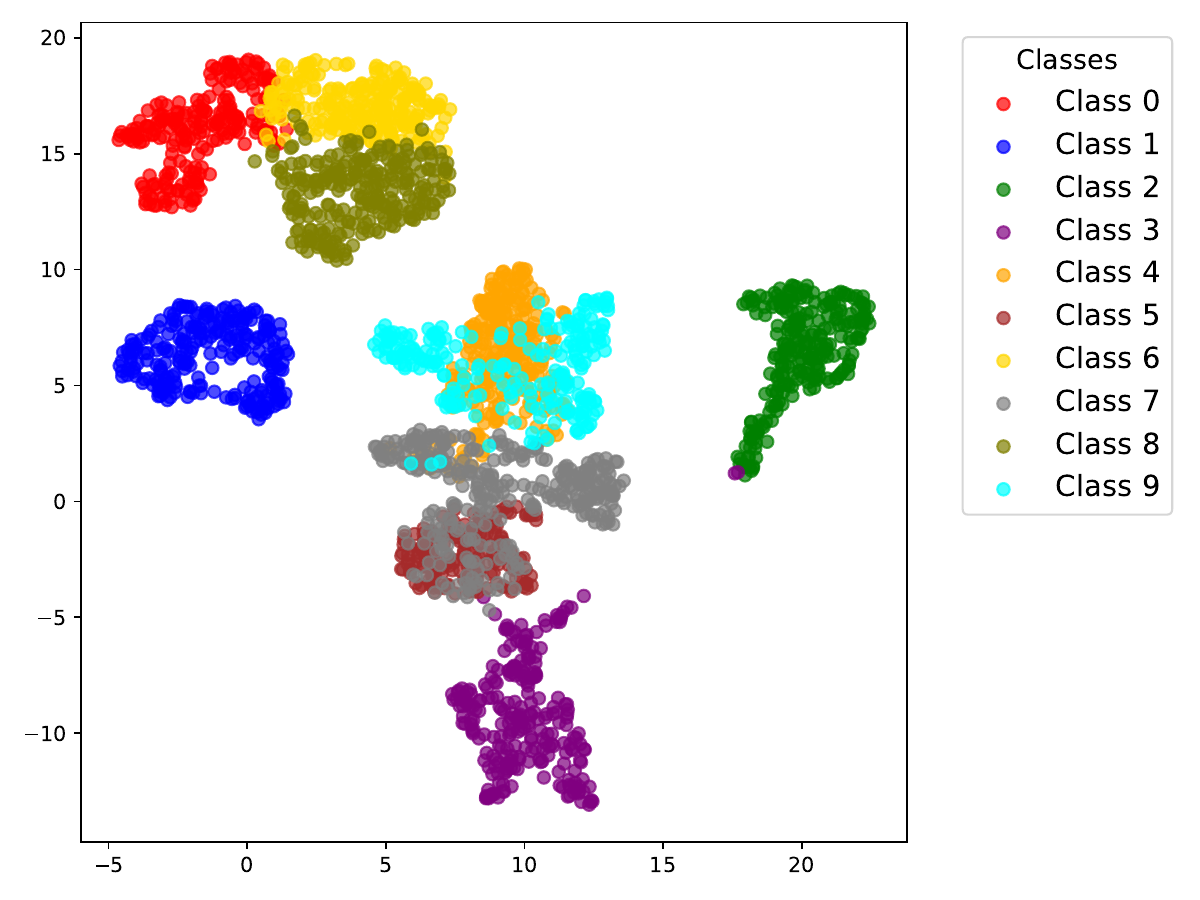}}
    \subfloat[\centering RNN ]{\includegraphics[scale=0.3, trim=0 0 120 0,clip]{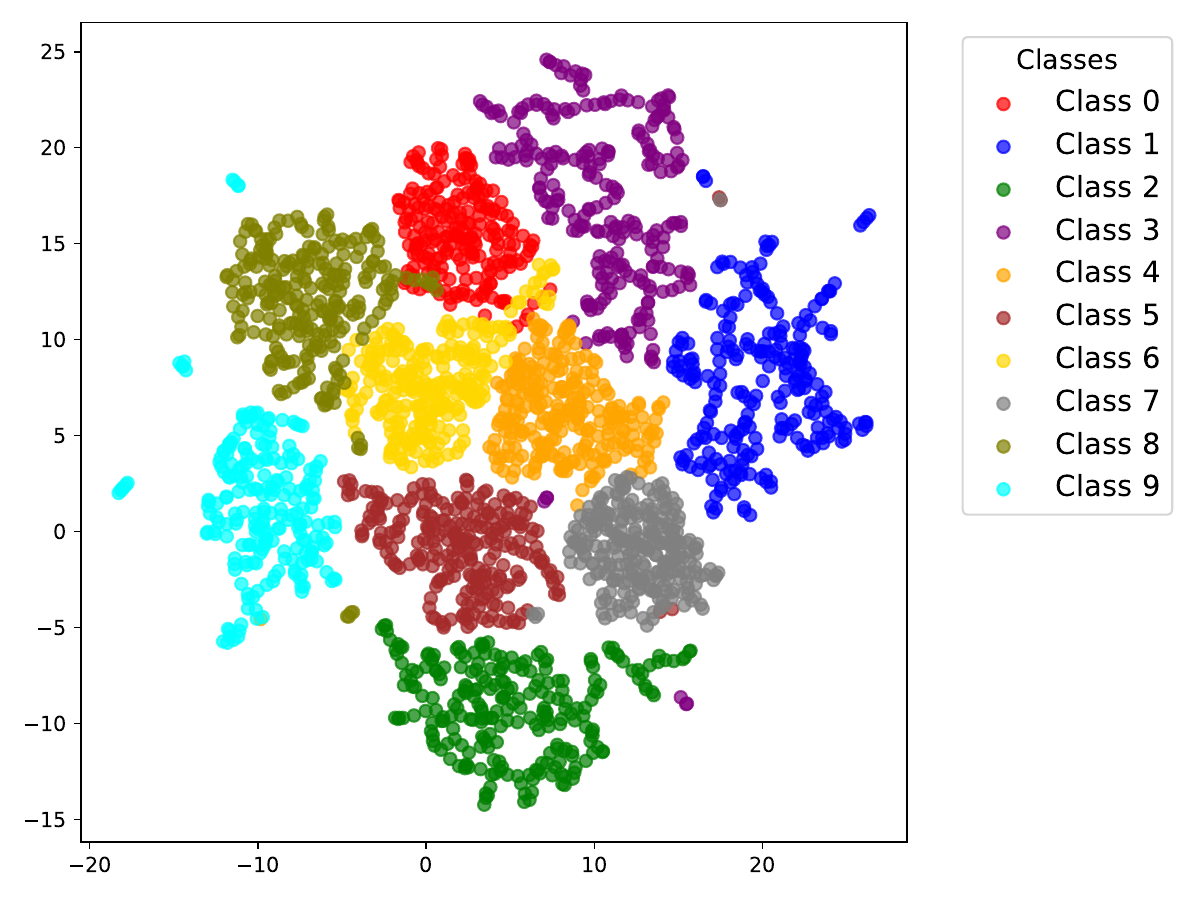}}
    \subfloat[\centering LSTM ]{\includegraphics[scale=0.3, trim=0 0 120 0,clip]{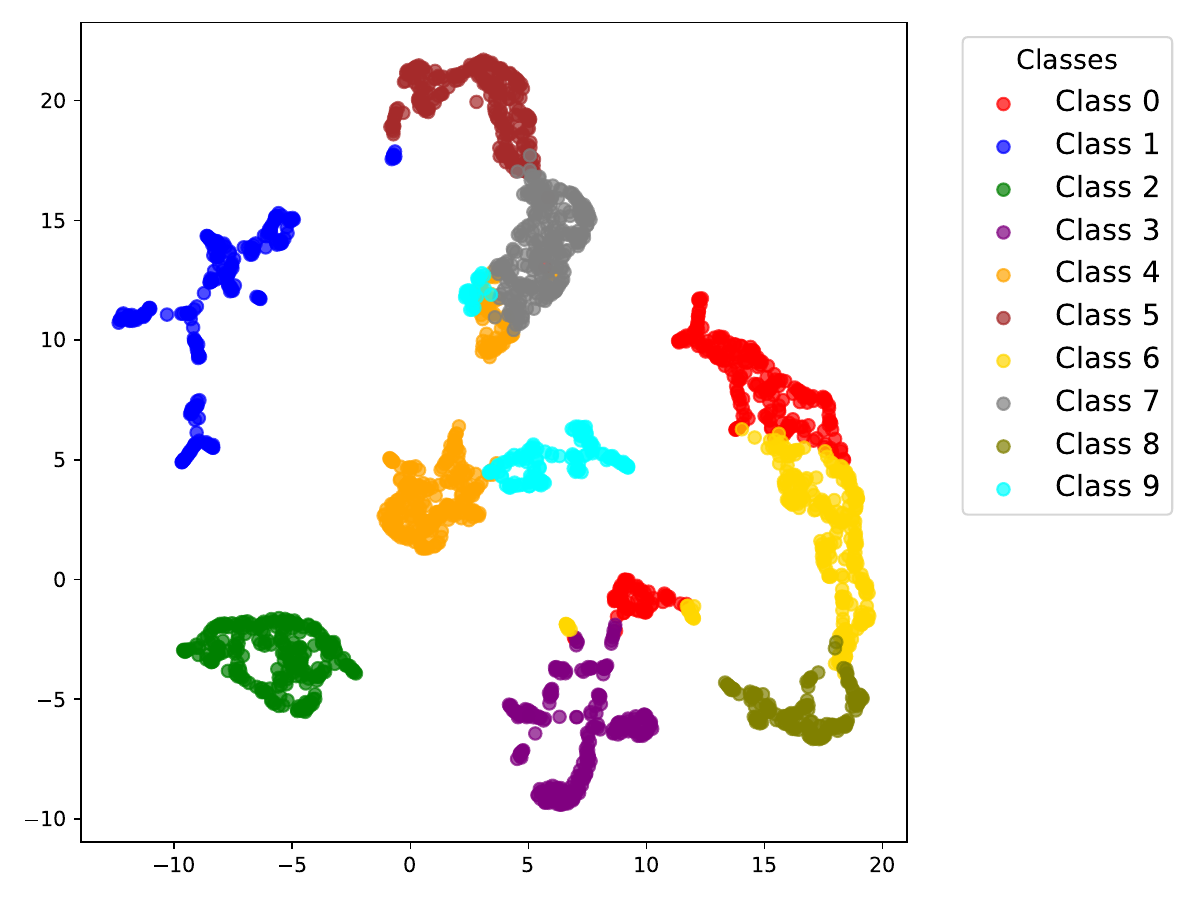}}
    \\
    \subfloat[\centering Transformer]{\includegraphics[scale=0.3, trim=0 0 120 0,clip]{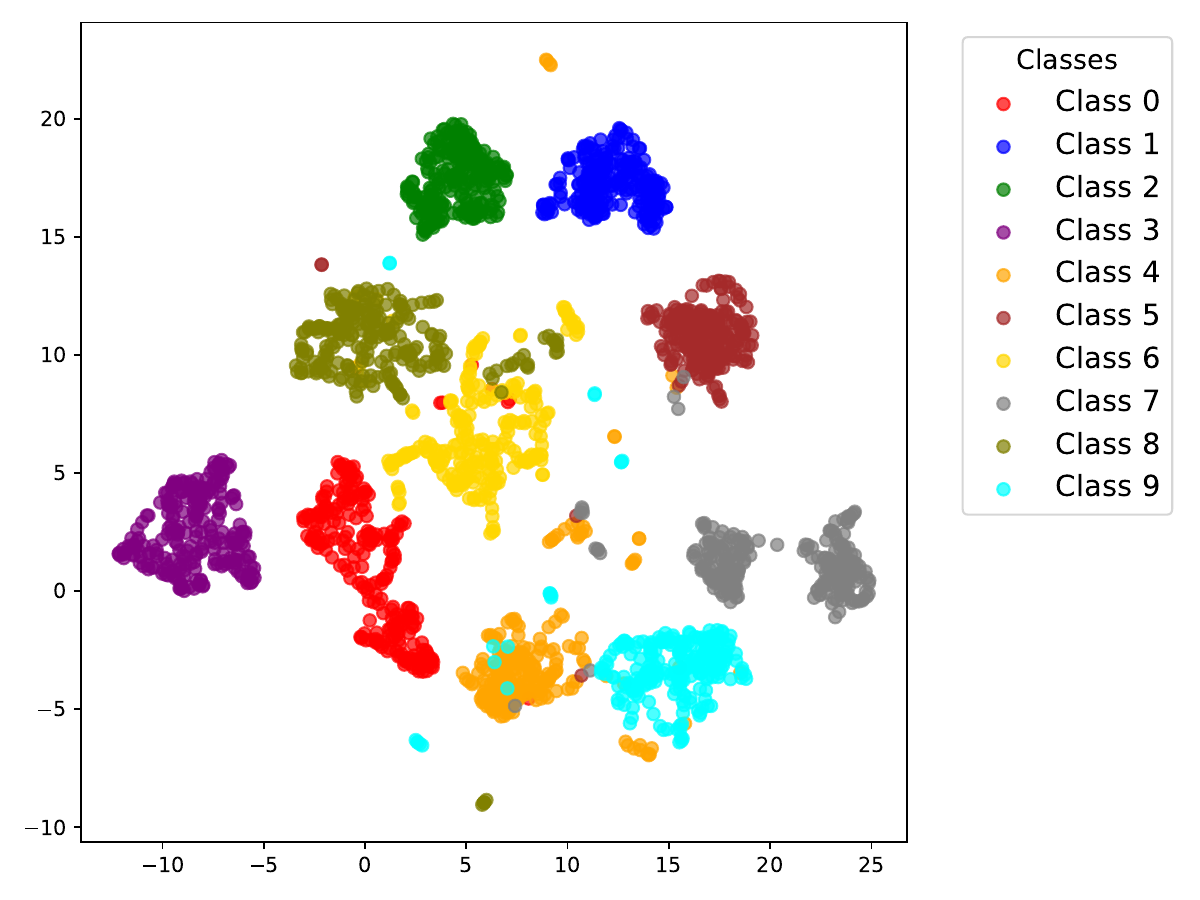}}
    \subfloat[\centering \textbf{Ours}]{\includegraphics[scale=0.3, trim=0 0 120 0,clip]{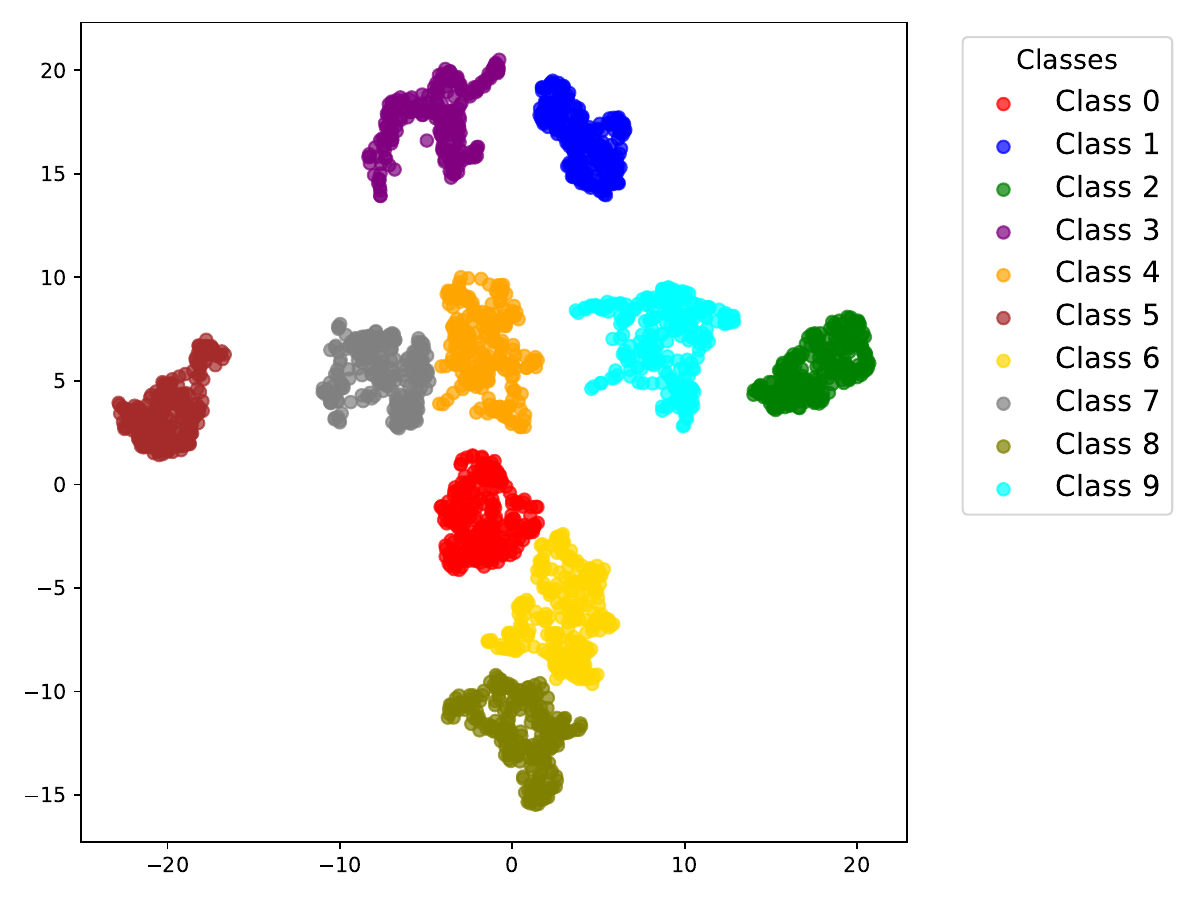}}
    \quad\quad\quad\quad
    \subfloat[\centering ]{\includegraphics[scale=0.45, trim=460 140 0 0,clip]{./projection_cnn.pdf}}
    \quad\quad\quad\quad\quad\quad
    \caption{The cluster analysis of the investigated networks.}
    \label{fig:umap}
\end{figure*}

Furthermore, a cluster analysis is conducted according to the uniform manifold approximation and projection (UMAP)~\cite{mcinnes2018umap}, to comprehensively validate the effectiveness of our DPFFN. 
Figure~\ref{fig:umap} gives the distribution of feature representations learned by the investigated networks. 
It can be seen that our DPFFN can reach a feature representation with distinctly compact and well-separated clusters. 
This suggests that the polarization features captured by our DPFFN are more discriminative and the class-specific information is more explicitly exhibitted. 
In contrast, other networks under investigation give overlapping and less cohesive clusters. 
The clustering results are consistent with the findings from the confusion matrices, further reinforce the effectiveness of our DPFFN in distinguishing between different classes.

\section{Discussion}\label{sec:discussion}

This section presents an in-depth investigation on our DPFFN, including the alblation study, the robustness of our DPFFN under noisy and incomplete data conditions, and the choice of hyperparameters. 

\subsection{Ablation Study}\label{sec:ablation_studies}
The corresponding results are summarized in Table \ref{tab:ablation}.
The network based on Transformer studied in Section~\ref{sec:capability_results} is presented as the baseline. 
In the study, Transformer network can be viewed as a starting point of our DPFFN. 
By gradual inclusion of the local subbranch, the feature fusion module and the fusion loss term, the roles these components played are individually revealed.
In particular, 
\begin{itemize}
    \item The impact of the local subbranch on the DPFFN can be revealed by the comparison between Transformer and the network with the local subbranch employed where only HH-polarized data is utilized.
    From Table~\ref{tab:ablation}, it can been seen that an accuracy increase from 88.4\% to 89.2\% due to the employment of the local subbranch is included. 
    \item The role of polarmetric scatterings is revealed by concatenating the HH- and VH-polarized features straightforwardly. 
    It can been seen that such a simple treatment improves the accuracy to 90.2\%, demonstrating the effectiveness of leveraging multiple polarizations for recognition tasks. 
    \item Our proposed feature fusion module can further enhance the accuracy to 92.0\%. Obviously, the proposed feature fusion module greatly improves the performance of the network by a nice understanding of the scattering. 
    \item The role the fusion loss term in Eq.~\eqref{eq:loss_function} played can be figured out by comparing the last two line of Table~\ref{tab:ablation}. Apparently, an improvement of 2.8\% can be reached by the inclusion of the fusion loss term. 
\end{itemize}

\begin{table}[htbp]
    \centering
    \caption{Ablation experiments results. Transformer network is used as the baseline.}
    \label{tab:ablation}
    \begin{tabular}{c|c}
        \hline
         & \textbf{Accuracy}\\
        \hline
        \cyan{Transformer (Baseline)} & \cyan{88.4\%} \\
        + Local Subbranch (HH Only) & 89.2\% \\
        \cyan{HH \& VH Concatenation} & \cyan{90.2\%} \\
        + Our Feature Fusion Module & 92.0\% \\
        \textbf{Fully Implemented DPFFN} & \textbf{94.8\%} \\
        \hline
    \end{tabular}
\end{table}

\begin{figure}[t]
    \centering
    \includegraphics[width=0.48\textwidth, trim = 0 5 0 27, clip]{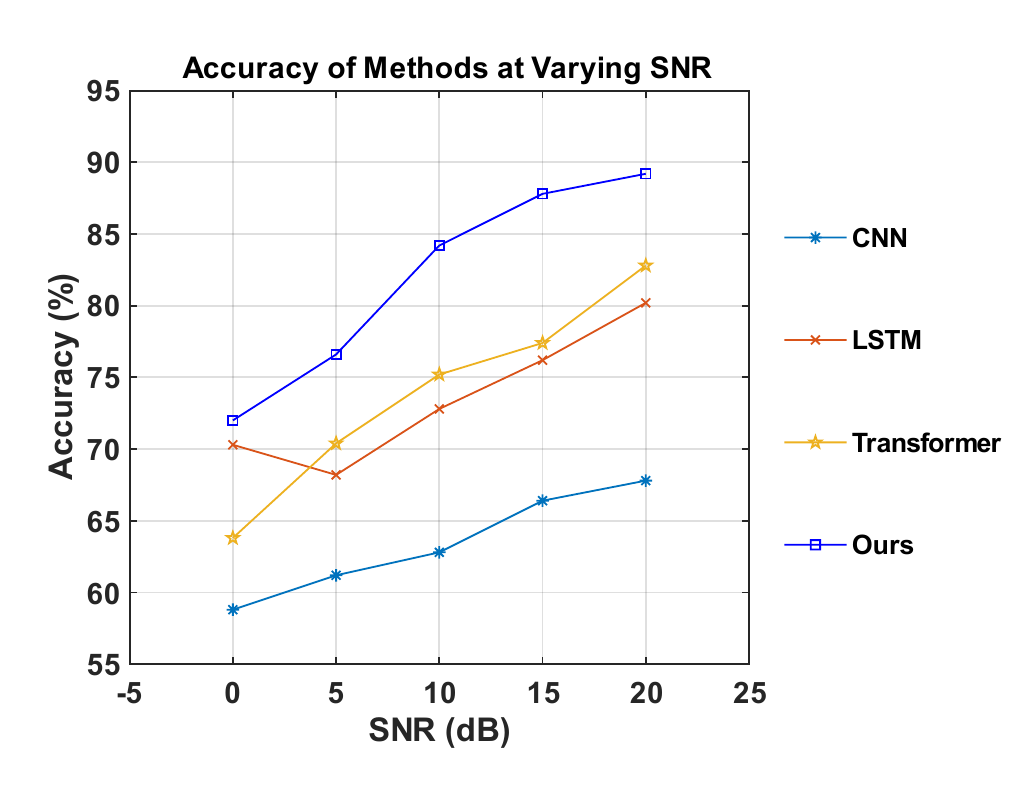}
    \caption{The classification accuracy of the investigated networks at varying SNR levels. }
    \label{fig:noise}
\end{figure}

\subsection{Robustness to Noise}
To assess the robustness of our DPFFN to noise, experiments are conducted with data corrupted by noise, simulating real-world conditions under various signal-to-noise ratio (SNR) scenarios. 
Specifically, Gaussian noise is added to the HRRP sequences at SNR levels ranging from 0 dB to 20 dB. 
The results of these experiments are presented in Fig. \ref{fig:noise}. 

It can been seen that our DPFFN is robust across varying SNR levels. 
Among networks utilizing data of the single HH-polarization, the network based on Transformer achieves the highest and most stable accuracy as SNR decreases.
This validates the choice of Transformer as the global module in our DPFFN. 
The CNN network seems robust to noise. 
The reason may lie in its inferior classification performance.

Interestingly, the LSTM network achieves a higher accuracy at the case with SNR = 0dB compared to that with SNR = 5dB. 
It is possibly due to overfitting to specific noise characteristics at SNR = 5dB, leading to a misalignment with the underlying signal patterns.

\subsection{Robustness to Missing Data}
In real-world scenarios, the integrity of HRRP data is often compromised due to factors such as signal attenuation, occlusions, and hardware limitations. 
To assess the robustness of our DPFFN under these conditions, experiments involving a random truncation of HRRP sequences are conducted. 

To simulate realistic data missing scenarios, specific percentages of HRRP sequences were randomly discarded. 
The recognition performance of our DPFFN is studied againest Transformer and LSTM networks. 
As summarized in Table \ref{tab:missing}, all investigated networks perform quite well. 
Our DPFFN demonstrates its robustness by delivering acceptable accuracies even with significant data missing, highlighting its resilience to incomplete input sequences. 
Specifically, when the missing rate increases to 50\%, 75\%, and 87.5\%, the accuracy of our DPFFN remained as high as 90.3\%, 88.0\%, and 85.7\%, respectively.

\begin{table}[htbp]
    \centering
    \caption{The classification accuracy at different missing data rates. Transformer and LSTM networks in Section~\ref{sec:capability_results} are included for comparison.}
    \label{tab:missing}
    \begin{tabular}{c|c|c|c}
        \hline
        \textbf{Data Missing Rate} & \cyanbf{Ours} & \textbf{Transformer} & \textbf{LSTM}\\
        \hline
        0\% & \textbf{94.8\%} & 88.4\% & 83.2\%\\
        50\% & \textbf{90.3\%} & 86.1\% & 77.6\%\\
        75\% & \textbf{88.0\%} & 82.6\% & 72.8\%\\
        87.5\% & \textbf{85.7\%} & 79.6\% & 67.9\%\\
        \hline
    \end{tabular}
\end{table}

\subsection{Hyerparameters}\label{sec:hyperparam}

Hyperparameters always play crucial roles in determining the performance of a network. 
There are several critical hyperparameters that may influence the performance of our DPFFN, such as, the number of heads in the multi-head attention mechanism, the number of transformer encoders in the global subbranch, and the number of stacked residual blocks in the local subbranch.
They are investigated in sufficient detail as shown in Table \ref{tab:hyperparam}. 

\begin{table}[htbp]
    \centering
    \caption{The impact of the hyperparameters on our DPFFN. Number of heads, number of global blocks, and number of local blocks are investigated. $M$ represents the number of transformer encoders in the global subbranch, while $N$ represents the number of stacked residual blocks in the local subbranch.}
    \label{tab:hyperparam}
    \resizebox{\linewidth}{!}{
    \begin{tabular}{c|c|c|c}
        \hline
        \textbf{Hyperparameter}&\textbf{Value} & \textbf{Accuracy} & \textbf{Params} \\
        \hline 
        \multirow{5}*{\makecell{\textbf{Number of Heads} \\ \textbf{($M = 10$ and $N = 10$)} }} 
        & 4 & 92.0\% &  13,595,326  \\
        ~&6 & 93.2\% &  14,627,518 \\
        ~&8 & 94.0\% &  15,659,710 \\
        ~&10 & 94.8\% & 16,691,902 \\
        ~&12 & 93.6\% & 17,724,094  \\
        \hline
        \multirow{5}*{\makecell{\textbf{Value of M} \\ \textbf{($heads = 10$ and $N = 10$)} }} 
        & 4 & 87.6\% & 12,378,958  \\
        ~&6 & 90.8\% & 13,816,606  \\
        ~&8 & 91.6\% & 15,254,254 \\
        ~&10 & 94.8\% & 16,691,902 \\
        ~&12 & 95.6\% & 18,129,550  \\
        \hline
        \multirow{5}*{\makecell{\textbf{Value of N} \\ \textbf{($heads = 10$ and $M = 10$)} }} 
        & 4 & 88.4\% & 16,691,614  \\
        ~&6 & 92.0\% & 16,691,710  \\
        ~&8 & 93.2\% & 16,691,806  \\
        ~&10 & 94.8\% & 16,691,902  \\
        ~&12 & 95.2\% & 16,691,998  \\
        \hline

    \end{tabular}}
\end{table}

It can be seen that the number of heads in the multi-head attention mechanism impacts the overall performance of our DPFFN slightly. 
Increasing the number of heads does not always obviously improve the accuracy. 
This phenomenon is consistent with the study in~\cite{head2019Michel}.
Our experiments here suggest that setting it equal to 8 can reach an optimal balance between accuracy and computational efficiency. 
Such a choice is preferred for our DPFFN. 
Table~\ref{tab:hyperparam} shows that increasing the number of global blocks can obviously enhance the classification accuracy. 
Unfortunately, the number of parameters increases quite fast with the number of global blocks. 
As a trade-off, 10 blocks are employed in our DPFFN. 

Similar to the global blocks, increasing the number of local blocks can also obviously enhance the classification accuracy. 
However, the number of parameters increases very slowly with the number of local blocks.
As a result, our DPFFN can choose a relatively larger number of local blocks.

\section{Conclusion}\label{sec:conclusion}

A dual-polarization feature fusion network (DPFFN) is proposed along with a novel two-stage feature fusion strategy for RATR based on multiple polarizations HRRP sequences. 
A specific fusion loss function is developed, enabling the adaptive generation of comprehensive multi-modal representations from polarimetric HRRP sequences.
Experimental results demonstrate that our DPFFN significantly outperforms several competitive networks, achieving a classification accuracy of 94.8\% on a simulated polarimetric HRRP dataset. 
The robustness of our DPFFN is also evaluated under noisy and incomplete data conditions, demonstrating its superior performance and resilience in challenging scenarios. 
The hyperparameter analysis further provides a guidance for an optimal configuration of the proposed network.

\bibliographystyle{IEEEtran}
\bibliography{IEEEabrv,reference}

\vfill

\end{document}